# Design-Based Estimation and Central Limit Theorems for Local Average Treatment Effects for RCTs

**November 2023**

Peter Z. Schochet, Ph.D. (Corresponding Author)
Senior Fellow, Associate Director
Mathematica
P.O. Box 2393
Princeton, NJ 08543-2393 USA
Phone: (609) 936-2783
pschochet@mathematica-mpr.com

## Abstract

There is a growing literature on design-based methods to estimate average treatment effects for randomized controlled trials (RCTs) using the underpinnings of experiments. In this article, we build on these methods to consider design-based regression estimators for the local average treatment effect (LATE) estimand for RCTs with treatment noncompliance. We prove new finite-population central limit theorems for a range of designs, including blocked and clustered RCTs, allowing for baseline covariates to improve precision. We discuss consistent variance estimators based on model residuals and conduct simulations that show the estimators yield confidence interval coverage near nominal levels. We demonstrate the methods using data from a private school voucher RCT in New York City USA.

# 1. Introduction

There is a growing literature on design-based methods to estimate average treatment effects (ATEs) for randomized controlled trials (RCTs). These methods use the building blocks of experimental designs to generate consistent, asymptotically normal ATE estimators with minimal assumptions. The underpinnings of these methods were introduced by Neyman (1923) and later developed in seminal works by Rubin (1974,1977) and Holland (1986) using a potential outcomes framework. The design-based approach is appealing in that it applies to continuous, binary, and discrete outcomes, allows for treatment effect heterogeneity, and makes no assumptions about the distribution of potential outcomes or the model functional form.

The design-based RCT literature has focused on ATE estimation for the intention-to-treat (ITT) estimand that pertains to the treatment offer. In this article, we build on these methods to develop design-based estimators for the local average treatment effect (LATE) estimand, also referred to as the complier average causal effect (CACE) estimand, that pertains to those who would receive the treatment if assigned to the treatment but not control condition.

The LATE estimand applies to RCTs where some in the treatment group may not receive the treatment and some in the control group might, and interest lies in estimating the mediating effects of treatment receipt on outcomes. A common example is a medical trial where only some treatment group members take the offered drug. Another example is an RCT testing an intervention (such as a text messaging campaign) encouraging the treatment group to participate in a program or activity available to both research groups, where interest lies in estimating the effects of actual participation on outcomes (Hummel and Maedche 2019; Mertens et al. 2022).

The ITT estimand is policy relevant when treatment group noncompliance is part of the natural study landscape (e.g., patients do not always take their medications). However, for RCTs where some controls receive the treatment or where compliance is artificially manipulated by the experiment (e.g., in random encouragement designs), ITT effects become more difficult to interpret. In these settings, the LATE estimand pertains to ''pure'' treatment effects—albeit for the unobserved subgroup of compliers—that can help guide future treatment improvement. Decision makers may also be interested in the LATE estimand if they believe treatment implementation can be improved to increase compliance.

In this article, we develop design-based, regression LATE estimators that derive from the Neyman-Rubin-Holland model that underlies experiments combined with standard monotonicity and exclusion restrictions found in the instrumental variables (IV) literature (Angrist et al., 1996; Imbens and Rubin, 2015). We prove new finite-population central limit theorems (CLTs) that build on the theory in Scott and Wu (1981), Imbens and Rubin (2015), Li and Ding (2017), Pashley (2019), and Schochet et al. (2022) for a range of RCT designs, including blocked and clustered RCTs that are common across fields (Bland, 2004; Schochet, 2008). We allow for baseline covariates to improve precision, allow for heterogenous treatment effects, and do not require that the true relationship between treatment receipt and outcomes is linear.

We discuss consistent variance estimators based on model residuals that are compared to standard IV estimators. Our simulations suggest that the design-based LATE estimators for the non-clustered RCT yield confidence interval coverage near nominal levels under the assumed identification conditions with modest sample sizes, although with some over-coverage. Finally, we apply the methods using data from the Evaluation of the New York City (NYC) School Choice Scholarships Program, an RCT conducted in the United States testing the effects of private school vouchers for low-income public school students in grades kindergarten to 4 (Mayer et al., 2002; Krueger and Zhu, 2004).

The rest of this article proceeds as follows. Section 2 discusses the related literature. Section 3 discusses the design-based framework, LATE estimation, and a CLT for the non-clustered RCT. Section 4 extends the results to blocked and clustered RCTs. Section 5 presents simulation results. Section 6 presents a case study using the NYC voucher study. Section 7 concludes. The online appendix provides proofs of main results and additional simulation details.

**2. Related literature**

Our work builds on the growing literature on design-based methods to estimate ATEs for full sample ITT analyses (Yang and Tsiatis, 2001; Freedman, 2008; Schochet, 2010, 2016; Aronow and Middleton, 2013; Lin, 2013; Imbens and Rubin, 2015; Middleton and Aronow, 2015; Li and Ding, 2017; Schochet et al., 2022). These methods have also been extended to estimate ATEs for discrete subgroups defined by pre-treatment sample and site characteristics (Schochet, 2023).

Our work also draws on the literature on IV estimation of treatment effects using a potential outcomes framework (e.g., Imbens and Angrist, 1994; Angrist et al., 1996; Abadie et al., 2002; Abadie, 2002, 2003; Imai et al., 2013; Imbens and Rubin, 2015; Hazard and Lowe, 2022). These articles, however, focus on a superpopulation framework where potential outcomes are assumed to be sampled from some broader universe. In contrast, we consider a finite-population RCT setting where potential outcomes are fixed, randomness is from treatment assignment only, and study results are assumed to pertain to the sample only—an often more realistic RCT scenario.

Of particular relevance to our work, Imbens and Angrist (1994) prove a superpopulation CLT for the LATE estimator for the non-clustered design without covariates. Our work instead uses a finite-population approach, where we (i) allow for covariates to improve precision, (ii) do not require a linear outcome-receipt model, (iii) extend our results for the non-clustered RCT to more general blocked and clustered RCTs, and (iv) examine design-based variance estimation.

Our work also builds on Li and Ding (2017) who consider design-based methods for the linear IV model (Imbens and Rosenbaum, 2005) where the ATE on response is modeled as a linear function of the ATE on treatment dosage with parameter $\tau$. Li and Ding (2017) back out asymptotic confidence intervals for $\tau$, but do not consider the asymptotic properties of the LATE estimator itself, which is our focus. Further, Li and Ding (2017) do not allow for covariates or extend their results to blocked and clustered RCTs, which are contributions of this work.

In what follows, we first discuss LATE estimation for the non-clustered RCT design to fix concepts. We then more briefly discuss extensions to blocked and clustered designs.

## 3. Design-based framework for LATE estimation for the non-clustered RCT

We assume an RCT of $n$ individuals, with $n^1 = np$ assigned to the treatment group and $n^0 = n(1-p)$ assigned to the control group, where $p$ is the treatment assignment rate ($0 < p < 1$). Let $T_i$ equal 1 if person $i$ is randomly assigned to the treatment group and 0 otherwise.

We conceptualize the potential treatment receipt decision, $D_i(T_i)$, as binary, and consider the general RCT setting where noncompliance can exist in both research conditions. Let $D_i(1) = 1$ if the person would receive the treatment if assigned to the treatment condition and 0 otherwise, and similarly for $D_i(0)$ in the control condition. We use similar notation to define potential outcomes, $Y_i(T_i, D_i(T_i))$, which can be continuous, binary, or discrete.

We assume a finite-population model, where potential outcomes are assumed fixed for the study, so that randomness and estimator variance arise due to $T_i$ only. Thus, study results are assumed to generalize to the study sample only rather than to a superpopulation often vaguely defined in the literature. The finite-population framework is often realistic for RCTs that usually rely on volunteer samples of persons and sites. The philosophy is that study results can be synthesized with other related information to assess intervention effects in broader settings and to inform policy decisions regarding wider implementation of the treatment.

Our analysis relies on two key assumptions, where others are discussed in subsequent sections. The first is the stable unit treatment value assumption (SUTVA) (Rubin, 1986; Angrist et al., 1996), and the second is complete randomization (Imbens and Rubin, 2015).

*(C1): SUTVA*: Let **T** be any vector of randomization realizations and let **D(T)** be the vector of treatment receipt decisions given **T**. Then, (i) if $T_i = T_i'$, then $D_i(\mathbf{T}) = D_i(\mathbf{T}')$; and (ii) if $T_i = T_i'$ and $D_i(\mathbf{T}) = D_i(\mathbf{T}')$ then $Y_i(\mathbf{T}, \mathbf{D}(\mathbf{T})) = Y_i(\mathbf{T}', \mathbf{D}'(\mathbf{T}'))$.

SUTVA implies that $D_i(T_i)$ and $Y_i(T_i, D_i(T_i))$ depend on the person's treatment assignment and not on those of others in the sample, and do not depend on the compliance of others. SUTVA also assumes a particular unit cannot receive different forms of the treatment.

*(C2): Complete randomization*: For fixed $n^1$, if $\mathbf{t} = (t_1, \ldots, t_n)$ is any vector of randomization realizations such that $\sum_{i=1}^{n} t_i = n^1$, then $Prob(\mathbf{T} = \mathbf{t}) = \binom{n}{n^1}^{-1}$.

This condition implies that each person has the same treatment assignment probability so that $T_i$ is independent of $D_i(1), D_i(0), Y_i(1, D_i(1))$, and $Y_i(0, D_i(0))$.

### 3.1. Estimands and identification

Under SUTVA, the ITT causal estimand on outcomes is the ATE in the finite population:

$$\tau_{ITT} = \bar{Y}(1) - \bar{Y}(0), \quad (1)$$

where $\bar{Y}(t) = \frac{1}{n}\sum_{i=1}^{n} Y_i(t, D_i(t))$ is the (fixed) mean potential outcome in research condition $t \in \{1,0\}$. Similarly, the ITT estimand for the mediating treatment receipt indicator is,

$$\pi_{ITT} = \bar{D}(1) - \bar{D}(0), \quad (2)$$

where $\bar{D}(t) = \frac{1}{n}\sum_{i=1}^{n} D_i(t)$ is the mean treatment receipt rate in research condition $t$.

The ITT estimand, $\tau_{ITT}$, combines the effects of the treatment on both recipients and nonrecipients in each research condition (compliers and noncompliers). To define the LATE estimand that pertains to compliers only, we follow Angrist et al. (1996) by defining principal strata based on the four possible $(D_i(1), D_i(0))$ combinations of receipt decisions made by the ITT population in the treatment and control conditions. The (1,1) stratum includes participants in both research conditions (always-takers), the (0,0) stratum includes non-participants in both conditions (never-takers), and the (1,0) and (0,1) strata include those who would participate in one research condition but not the other (compliers and defiers). Without further conditions, we cannot determine the specific stratum for any individual, because treatment receipt decisions are observed in one research condition only.

Using this framework, the ATE for those in principal stratum $(d, d')$ is,

$$\tau_{dd'} = \bar{Y}_{dd'}(1) - \bar{Y}_{dd'}(0) \tag{3}$$

for $d, d' \in \{1,0\}$. Thus, $\tau_{ITT}$ in (1) is a weighted average of the four $\tau_{dd'}$ estimands in (3):

$$\tau_{ITT} = p_{11}\tau_{11} + p_{10}\tau_{10} + p_{01}\tau_{01} + p_{00}\tau_{00}, \tag{4}$$

where $p_{dd'}$ are fixed stratum shares that sum to 1.

The LATE estimand is $\tau_{10}$, which pertains to the ATE for compliers. To identify this estimand, we invoke the following additional assumptions:

*(C3): Monotonicity*: $D_i(1) \geq D_i(0)$, which means that assignment to the treatment group can only increase intervention receipt, so that there are no defiers in the ITT population ($p_{01} = 0$). This condition rules out situations, for instance, in a random encouragement design where the encouragement conveys information about participation in a program or activity that some individuals may deem undesirable.

Under monotonicity, $\bar{D}(1) = p_{11} + p_{10}$ and $\bar{D}(0) = p_{11}$. Thus, solving for $p_{10}$ yields,

$$p_{10} = \pi_{ITT} = \bar{D}(1) - \bar{D}(0), \tag{5}$$

which is the ATE on treatment receipt.

*(C4): Exclusion restriction*: $Y_i(1,d) = Y_i(0,d)$ for $d \in \{1,0\}$, which means that conditional on the treatment receipt decision, the same outcome would result in the treatment or control condition. This implies that any effect of the treatment on longer-term outcomes must be via an effect on treatment receipt (the mediating mechanism). This restriction implies that the treatment has no effect on the always-takers and never-takers ($\tau_{11} = 0$ and $\tau_{00} = 0$). It also implies that we can write $Y_i(T_i) = Y_i(T_i, D_i(T_i))$, which helps reduce notation.

*(C5): Presence of compliers*: $0 < p_{10} < 1$, which means that being assigned to the treatment group will incentivize some sample members to receive the treatment who otherwise would not. Note that if $p_{10} = 1$, the LATE and ITT estimands align, so we do not consider this case.

Under *(C1)* to *(C5)*, the LATE estimand can then be identified from (4) as follows:

$$\tau_{10} = \frac{\tau_{ITT}}{p_{10}} = \frac{\tau_{ITT}}{\pi_{ITT}}, \tag{6}$$

where the second equality uses (5). Intuitively, the LATE estimand inflates the ITT estimand to pertain to compliers only.

### 3.2. Estimators

We can estimate $\tau_{10}$ in (6) using $\hat{\tau}_{10} = \hat{\tau}_{ITT}/\hat{\pi}_{ITT}$, where $\hat{\tau}_{ITT}$ and $\hat{\pi}_{ITT}$ are consistent estimators of $\tau_{ITT}$ and $\pi_{ITT}$. We first consider design-based estimation of $\hat{\tau}_{ITT}$. To do this, note that under the potential outcomes framework and SUTVA, the data generating process for the observed outcome, $y_i = Y_i(T_i)$, is as follows:

$$y_i = T_i Y_i(1) + (1 - T_i) Y_i(0). \tag{7}$$

This relation states that we can observe $y_i = Y_i(1)$ for those in the treatment group and $y_i = Y_i(0)$ for those in the control group, but not both.

Next, rearranging (7) generates the following regression model that underlies experiments:

$$y_i = \alpha + \tau_{ITT}\tilde{T}_i + u_i, \tag{8}$$

where $\tilde{T}_i = (T_i - p)$ is the centered treatment indicator; $\alpha = p\bar{Y}(1) + (1-p)\bar{Y}(0)$ is the intercept (expected potential outcome); $\tau_{ITT} = \bar{Y}(1) - \bar{Y}(0)$; and the "error" term, $u_i$, is,

$$u_i = T_i(Y_i(1) - \bar{Y}(1)) + (1 - T_i)(Y_i(0) - \bar{Y}(0)).$$

We center the treatment indicator in (8) to facilitate the theory without changing the estimators.

In contrast to usual formulations of the regression model, the residual, $u_i$, is random solely due to $T_i$ (Freedman, 2008; Schochet, 2010; Lin, 2013). This framework allows individual-level treatment effects, $\tau_i = Y_i(1) - Y_i(0)$, to vary across the sample, and is nonparametric because it makes no assumptions about the potential outcome distributions. The model does not satisfy key assumptions of the usual regression model: over the randomization distribution ($R$), $u_i$ is heteroscedastic, and $E_R(u_i)$, $Cov_R(u_i, u_{i'})$, and $E_R(\tilde{T}_i u_i)$ are nonzero if $\tau_i$ varies across $i$.

Consider ordinary least squares (OLS) estimation of a working model that includes in (8) a $1xV$ vector of fixed, centered baseline covariates, $\tilde{\mathbf{x}}_i = (\mathbf{x}_i - \bar{\mathbf{x}})$, with parameter vector, $\boldsymbol{\beta}$:

$$y_i = \alpha + \tau_{ITT}\tilde{T}_i + \tilde{\mathbf{x}}_i\boldsymbol{\beta} + e_i, \tag{9}$$

where $\bar{\mathbf{x}} = n^{-1}\sum_{i=1}^{n}\mathbf{x}_i$ are covariate means and $e_i$ is the error term. While the covariates do not enter the true RCT model in (8) and the ATE estimands do not change, they will increase precision to the extent they are correlated with the potential outcomes. We do not need to assume that the true conditional distribution of $y_i$ given $\mathbf{x}_i$ is linear in $\mathbf{x}_i$. We define $\boldsymbol{\beta}$ in Section 3.3.

OLS estimation of (9) yields the following covariate-adjusted estimator for $\tau_{ITT}$:

$$\hat{\tau}_{ITT} = (\bar{y}^1 - \bar{y}^0) - (\bar{\mathbf{x}}^1 - \bar{\mathbf{x}}^0)\hat{\boldsymbol{\beta}}, \tag{10}$$

where $\bar{y}^t = \frac{1}{n^t}\sum_{i:T_i=t}^n y_i$ and $\bar{\mathbf{x}}^t = \frac{1}{n^t}\sum_{i:T_i=t}^n \mathbf{x}_i$ for $t \in \{1,0\}$ are observed means for the two groups, and $\hat{\boldsymbol{\beta}}$ is the OLS estimator for $\boldsymbol{\beta}$ (Li and Ding 2017; Schochet et al. 2022). Intuitively, (10) adjusts the differences-in-means estimator by the covariates. Note that the same estimator results using a non-centered model in (9) using $T_i$ and $\mathbf{x}_i$. Our results can also be extended to models which interact $\tilde{T}_i$ and $\tilde{\mathbf{x}}_i$, but the pooled model is typically used in practice.

The same design-based approach can be used to obtain the covariate-adjusted ITT estimator for treatment receipt, $\hat{\pi}_{ITT}$. To do this, we replace $y_i$ in (9) with the observed treatment receipt indicator, $d_i = D_i(T_i)$, assuming the same covariates as in (9) (as is typical in practice) with parameter vector, $\boldsymbol{\gamma}$. This yields the following OLS estimator:

$$\hat{\pi}_{ITT} = (\bar{d}^1 - \bar{d}^0) - (\bar{\mathbf{x}}^1 - \bar{\mathbf{x}}^0)\hat{\boldsymbol{\gamma}}, \tag{11}$$

where $\bar{d}^t = \frac{1}{n^t}\sum_{i:T_i=t}^n d_i$ is the treatment receipt rate for group $t$ and $\hat{\boldsymbol{\gamma}}$ is the OLS estimator for $\boldsymbol{\gamma}$.

Using the ITT estimators in (10) and (11), we can now obtain the following consistent ratio estimator for the LATE estimand (see Theorem 1 in Section 3.3):

$$\hat{\tau}_{10} = \frac{\hat{\tau}_{ITT}}{\hat{\pi}_{ITT}} = \frac{(\bar{y}^1 - \bar{y}^0) - (\bar{\mathbf{x}}^1 - \bar{\mathbf{x}}^0)\hat{\boldsymbol{\beta}}}{(\bar{d}^1 - \bar{d}^0) - (\bar{\mathbf{x}}^1 - \bar{\mathbf{x}}^0)\hat{\boldsymbol{\gamma}}}. \tag{12}$$

In the RCT setting, the LATE estimand with covariates can be interpreted as the ATE among complier groups with weights proportional to the probability of compliance given the covariates (Angrist and Imbens, 1995; Blandhol et al., 2022).

Mechanically, $\hat{\tau}_{10}$ is an IV estimator using $\mathbf{z}_i = (1\ T_i\ \mathbf{x}_i)$ as the instrument vector for a regression of $y_i$ on $\mathbf{d}_i^* = (1\ d_i\ \mathbf{x}_i)$ (i.e., $\hat{\tau}_{10} = [\sum_{i=1}^n (\mathbf{z}_i'\mathbf{d}_i^*)^{-1}\mathbf{z}_i'y_i]_{(2,1)}$). It is also a two-stage least squares (2SLS) estimator where $d_i$ is regressed on $\mathbf{z}_i$ in the first stage, and $y_i$ is regressed on $\hat{\mathbf{d}}_i^* = (1\ \hat{d}_i\ \mathbf{x}_i)$ in the second stage, where $\hat{d}_i$ are first-stage predicted values. The resulting 2SLS parameter estimate on $\hat{d}_i$ yields (12). However, the usual IV framework differs from ours as it assumes a superpopulation model, a linear outcome-receipt model, and constant treatment effects (though Imbens and Rubin, 2015 mention these assumptions are not needed), leading to differences in the asymptotic theory.

In finite samples, if $T_i$ is a weak instrument for $d_i$, then $\hat{\tau}_{10}$ will be biased towards the least squares estimator in the second stage outcome model (Stock and Yogo, 2005). Although we consider large-sample properties of $\hat{\tau}_{10}$, examining instrument strength is important in practice, which in our setting could be assessed, for example, by examining whether the F-statistic from a regression of $d_i$ on $T_i$ exceeds 16 (Stock and Yogo, 2005; Table 2).

### 3.3. Asymptotic results

To consider the asymptotic properties of $\hat{\tau}_{10}$ in (12), we consider a hypothetical increasing sequence of finite populations where $n \to \infty$. Parameters should be subscripted by $n$, but we omit this notation for simplicity. We assume that $n^1/n \to p^*$ as $n \to \infty$, so that both $n^1$ and $n^0$ increase with $n$. In addition, we assume that $p_{10} \to p_{10}^*$ where $0 < p_{10}^* < 1$, so that the number of compliers also grows with $n$, and similarly for $p_{00}$ and $p_{11}$, where $p_{10}^* + p_{00}^* + p_{11}^* = 1$.

Our new finite-population CLT for $\hat{\tau}_{10}$ builds on results in Scott and Wu (1981), Li and Ding (2017), and Schochet et al. (2022) who provide finite-population CLTs for ATE estimators, but not for LATE estimators. Before presenting our CLT, we need to define several terms. First, for $t \in \{1,0\}$, let $R_i(t) = \frac{1}{\pi_{ITT}}(\tilde{Y}_i(t) - \tilde{\mathbf{x}}_i \boldsymbol{\beta} - \tau_{10}(\tilde{D}_i(t) - \tilde{\mathbf{x}}_i \boldsymbol{\gamma}))$ denote linearized model residuals, where $\tilde{Y}_i(t) = Y_i(t) - \bar{Y}(t)$ and $\tilde{D}_i(t) = D_i(t) - \bar{D}(t)$ are centered variables. Second, let $S_R^2(t) = \frac{1}{n-1}\sum_{i=1}^n R_i^2(t)$ denote the variance of $R_i(t)$, and let $S_R^2(1,0) = \frac{1}{n-1}\sum_{i=1}^n R_i(1)R_i(0)$ denote the treatment-control covariance. Third, we define $\bar{Q} = \frac{1}{n^1}\sum_{i:T_i=1}^{n} R_i(1) - \frac{1}{n^0}\sum_{i:T_i=0}^{n} R_i(0)$ as the mean treatment-control difference in the $R_i(t)$ residuals, with variance:

$$\text{Var}(\bar{Q}) = \frac{S_R^2(1)}{n^1} + \frac{S_R^2(0)}{n^0} - \frac{S^2(\tau_{10})}{n}, \tag{13}$$

where $S^2(\tau_{10}) = \frac{1}{n-1}\sum_{i=1}^{n}(R_i(1) - R_i(0))^2$ is the heterogeneity of the LATE effects. Fourth, we define variances for the outcome, $S_Y^2(t) = \frac{1}{(n-1)\pi_{ITT}^2}\sum_{i=1}^{n}\tilde{Y}_i^2(t)$; for each covariate $v \in \{1,\ldots,V\}$, $S_{x,v}^2 = \frac{1}{(n-1)\pi_{ITT}^2}\sum_{i=1}^{n}\tilde{x}_{i,v}^2$; and for the full covariate set, $\mathbf{S}_\mathbf{x}^2 = \frac{1}{n}\sum_{i=1}^{n}\tilde{\mathbf{x}}_i'\tilde{\mathbf{x}}_i$. Finally, we define covariances between $\tilde{Y}_i(t)$, $\tilde{\mathbf{x}}_i$, and $D_i(t)$: $\mathbf{S}_{\mathbf{x},Y}^2(t) = \frac{1}{n}\sum_{i=1}^{n}\tilde{\mathbf{x}}_i'Y_i(t)$; $\mathbf{S}_{\mathbf{x}Y}^2(t) = \frac{1}{n}\sum_{i=1}^{n}(\tilde{\mathbf{x}}_i'Y_i(t) - \bar{\boldsymbol{\theta}})^2$, where $\bar{\boldsymbol{\theta}} = \frac{1}{n}\sum_{i=1}^{n}\tilde{\mathbf{x}}_i'Y_i(t)$; and similarly for $\mathbf{S}_{\mathbf{x},D}^2(t)$, $\mathbf{S}_{\mathbf{x}D}^2(t)$, $S_{Y,D}^2(t)$, and $S_{YD}^2(t)$.

We now present our CLT theorem, proved in Appendix A.

**Theorem 1.** Assume (*C1*) to (*C5*), and the following added conditions for $t \in \{1,0\}$:

(C6) Letting $g(t) = \max_{1\le i\le n}\{R_i^2(t)\}$, as $n \to \infty$,

$$\frac{1}{(n^t)^2}\frac{g(t)}{\text{Var}(\bar{Q})} \to 0.$$

(C7) $f^1 = n^1/n$ and $f^0 = n^0/n$ have limiting values, $p^*$ and $(1-p^*)$, for $0 < p^* < 1$.

(C8) The complier share, $p_{10} = \pi_{ITT}$, converges to $p_{10}^*$, for $0 < p_{10}^* < 1$; and similarly for $p_{00}$ and $p_{11}$, where $(p_{10}^* + p_{00}^* + p_{11}^*) = 1$.

(C9) Letting $h_v(t) = \max_{1\le i\le n}\{\tilde{x}_{i,v}^2\}$ for all $v \in \{1,\ldots,V\}$, as $n \to \infty$,

$$\frac{1}{\min(n^1,n^0)}\frac{h_v(t)}{S_{x,v}^2} \to 0.$$

(C10) $S_R^2(t)$, $S_R^2(1,0)$, $S_Y^2$, $S_{x,v}^2$, $\mathbf{S}_\mathbf{x}^2$, and $\mathbf{S}_{\mathbf{x},Y}^2(t)$ have finite limits.

Then, as $n \to \infty$, the LATE estimator in (12), $\hat{\tau}_{10}$, is a consistent estimator for $\tau_{10}$, and

$$\frac{\hat{\tau}_{10} - \tau_{10}}{\sqrt{\text{Var}(\bar{Q})}} \xrightarrow{d} N(0,1),$$

where $\text{Var}(\bar{Q})$ is defined as in (13).

*Remark 1*. (*C6*) and (*C9*) are Lindeberg-type conditions from Li and Ding (2017) that control the tails of the potential outcome and covariate distributions.

*Remark 2*. The first two terms for $\text{Var}(\bar{Q})$ in (13) pertain to separate variances for the two research groups because we allow for heterogeneous effects. The third term pertains to the treatment-control covariance, $S_R^2(1,0)$, expressed in terms of the heterogeneity of effects, $S^2(\tau_{10})$, which cannot be identified from the data but can be bounded (Aronow et al., 2014).

*Remark 3*. With constant treatment effects, as often assumed for IV estimators produced by standard statistical packages (such as Stata, R and SAS), we have that $S_R^2(1) = S_R^2(0)$ and $S^2(\tau_{10}) = 0$, so (13) reduces to, $\text{Var}(\bar{Q}) = \left(\frac{1}{n^1} + \frac{1}{n^0}\right)\left(\frac{n-1}{n-2}\right)(S_R^2(1) + S_R^2(0))$. Section 3.4 compares this IV variance to the design-based variance in (13) in more detail.

*Remark 4*. To help interpret $\text{Var}(\bar{Q})$, it is useful to expand $S_R^2(t)$ into its component parts:

$$S_R^2(t) = S_{RY}^2(t) + S_{RD}^2(t) + S_{RYD}^2(t)$$

$$= \frac{1}{\pi_{ITT}^2(n-1)}\sum_{i=1}^n (\tilde{Y}_i(t) - \tilde{\mathbf{x}}_i\boldsymbol{\beta})^2 + \frac{\tau_{10}^2 n}{\pi_{ITT}^2(n-1)}\sum_{i=1}^n (\tilde{D}_i(t) - \tilde{\mathbf{x}}_i\boldsymbol{\gamma})^2 \qquad (14)$$

$$- \frac{2\tau_{10}}{\pi_{ITT}^2(n-1)}\sum_{i=1}^n (\tilde{Y}_i(t) - \tilde{\mathbf{x}}_i\boldsymbol{\beta})(\tilde{D}_i(t) - \tilde{\mathbf{x}}_i\boldsymbol{\gamma}).$$

The first term, $S_{RY}^2(t)$, captures estimation error in $\hat{\tau}_{ITT}$ assuming treatment receipt rates are known, while the second and third terms are "corrections". The $S_{RD}^2(t)$ term captures estimation error in the treatment receipt rates and is positive, and $S_{RYD}^2(t)$ captures the outcome-receipt covariance that can be positive or negative. Thus, the total effect of the corrections on variance is indeterminate, as are the relative sizes of the z-scores for $\hat{\tau}_{10}$ and $\hat{\tau}_{ITT}$.

A key feature of (14) is that standard errors increase proportionally with $\pi_{ITT}$. Thus, precision can be a concern if ITT effects on treatment receipt are small (which leads to the weak instrument problem in finite samples). Another key feature is that the $S_{RY}^2(t)$ term is likely to be the main source of variance for typical RCTs. For instance, examining data from 10 large RCTs, Schochet and Chiang (2011) find that the two correction terms change the standard errors by a mean of only 0.20 percent (median of 0.06 percent), with no consistent sign for the covariances.

*Remark 5.* The CLT in Theorem 1 is proved by expressing $(\hat{\tau}_{10} - \tau_{10})/\sqrt{\text{Var}(\bar{Q})}$ as,

$$\frac{(\hat{\tau}_{10} - \tau_{10})}{\sqrt{\text{Var}(\bar{Q})}} = \left[\frac{1}{\sqrt{\text{Var}(\bar{Q})}} \frac{(\hat{\tau}_{ITT} - \tau_{10}\hat{\pi}_{ITT})}{\pi_{ITT}}\right]\left[\frac{\pi_{ITT}}{\hat{\pi}_{ITT}}\right]. \tag{15}$$

As shown in Appendix A, under the conditions in Theorem 1, we can adapt Theorem 4 in Li and Ding (2017) to obtain a CLT for the first bracketed term in (15), and the second bracketed term converges in probability to 1. Thus, the CLT in Theorem 1 follows by Slutsky's theorem.

*Remark 6.* As shown in Appendix A, the covariate parameter vector in (9) is $\boldsymbol{\beta} = (\mathbf{S}_\mathbf{x}^2)^{-1}[p\mathbf{S}_{\mathbf{x},Y}^2(1) + (1-p)\mathbf{S}_{\mathbf{x},Y}^2(0)]$, which is the (hypothetical) OLS coefficient that would result from a regression of $[pY_i(1) + (1-p)Y_i(0)]$ on $\tilde{\mathbf{x}}_i$. Similarly, in the treatment receipt model, we have that $\boldsymbol{\gamma} = (\mathbf{S}_\mathbf{x}^2)^{-1}[p\mathbf{S}_{\mathbf{x},D}^2(1) + (1-p)\mathbf{S}_{\mathbf{x},D}^2(0)]$.

*Remark 7.* Theorem 1 also applies to models without covariates by setting $\boldsymbol{\beta} = \mathbf{0}$ and/or $\boldsymbol{\gamma} = \mathbf{0}$. The covariates can also differ in the two models, which no longer yields an IV estimator.

### 3.4. Variance estimation

A simple plug-in (upper bound) variance estimator for (13) based on estimated, linearized OLS regression residuals from the outcome and treatment receipt models is as follows:

$$\hat{\text{Var}}(\bar{Q}) = \frac{s_R^2(1)}{n^1} + \frac{s_R^2(0)}{n^0}, \tag{16}$$

where

$$s_R^2(t) = \frac{1}{\hat{\pi}_{ITT}^2(n^t - k^t - 1)} \sum_{i:T_i=t}^{n} (\tilde{y}_i^t - \tilde{\mathbf{x}}_i^t\widehat{\boldsymbol{\beta}} - \hat{\tau}_{10}(\tilde{d}_i^t - \tilde{\mathbf{x}}_i^t\widehat{\boldsymbol{\gamma}}))^2 = \frac{1}{\hat{\pi}_{ITT}^2(n^t - k^t - 1)}RSS(t)$$

are mean-squared residuals for $t \in \{1,0\}$; $\tilde{y}_i^t = y_i - \bar{y}^t$, $\tilde{d}_i^t = d_i - \bar{d}^t$, and $\tilde{\mathbf{x}}_i^t = \mathbf{x}_i - \bar{\mathbf{x}}^t$ are group-centered variables; $k^1 = Vp$ and $k^0 = V(1-p)$ are degrees of freedom (*df*) adjustments for the covariates; and $\hat{\pi}_{ITT}$ and $\hat{\tau}_{10}$ are defined in (10) and (11). Note that $(\tilde{y}_i^t - \tilde{\mathbf{x}}_i^t\widehat{\boldsymbol{\beta}}) = (y_i - \hat{\alpha} - \hat{\tau}_{ITT}(t-p) - \tilde{\mathbf{x}}_i\widehat{\boldsymbol{\beta}})$ using (9), and similarly for $(\tilde{d}_i^t - \tilde{\mathbf{x}}_i^t\widehat{\boldsymbol{\gamma}})$. Here, the *df* losses due to the $V$ covariates are split across the two groups. Hypothesis testing can be conducted using t-tests with $df = (n - V - 2)$ or z-tests. Appendix B proves the following consistency result for $s_R^2(t)$:

**Theorem 2.** Under the conditions of Theorem 1, $s_R^2(t) - S_R^2(t) \xrightarrow{p} 0$ as $n \to \infty$ for $t \in \{1,0\}$.

Finally, we can compare (16) to the IV variance estimator with constant treatment effects produced by common statistical packages. To do this, consider the model without covariates. Then, the IV variance is, $\text{V}\hat{\text{a}}\text{r}(\bar{Q}_{IV}) = s^2(\frac{1}{n^1} + \frac{1}{n^0})$ with $s^2 = \frac{RSS(1)+RSS(0)}{\hat{\pi}_{ITT}^2(n-2)}$, or equivalently,

$\text{V}\hat{\text{a}}\text{r}(\bar{Q}_{IV}) = \frac{n}{(n-2)}[\frac{(n^1-1)}{n^0}\frac{s_R^2(1)}{n^1} + \frac{(n^0-1)}{n^1}\frac{s_R^2(0)}{n^0}]$. Ignoring the *df* corrections, we see that the sign of $[\text{V}\hat{\text{a}}\text{r}(\bar{Q}_{IV}) - \text{V}\hat{\text{a}}\text{r}(\bar{Q})]$ depends on the sign of $(2p-1)[\frac{s_R^2(1)}{n^1} - \frac{s_R^2(0)}{n^0}]$. If $p = .5$, the two variances are equal. If $\frac{s_R^2(1)}{n^1} > \frac{s_R^2(0)}{n^0}$, the likely case with heterogeneity, the IV variance is larger if $p > .5$ but not if $p < .5$. Thus, the relative sizes of the two variances will depend on $p$ and the heterogeneity pattern. A similar pattern holds for the model with covariates.

## 4. Extensions to blocked and clustered designs

### *4.1. Blocked RCTs*

In blocked designs, the sample is first divided into strata (e.g., sites, demographic groups, or time cohorts), and a mini-experiment is conducted in each one. To consider this design, we use similar notation as above with the addition of the subscript $b = (1, 2, \ldots, h)$ to indicate blocks. For instance, $T_{ib}$ is the treatment indicator in the block, $p_b$ is the block treatment assignment rate, $p_{10,b}$ is the complier sample share, and so on. Further, we define $S_{ib}$ as a 1/0 indicator of block membership, and $q_b = n_b/n$ as the block share.

The LATE estimand in block $b$ is, $\tau_{10,b} = \tau_{ITT,b}/\pi_{ITT,b}$, where $\tau_{ITT,b} = \bar{Y}_b(1) - \bar{Y}_b(0)$ and $\pi_{ITT,b} = \bar{D}_b(1) - \bar{D}_b(0)$ are block-specific ITT estimands. The pooled estimand across blocks is,

$$\tau_{10,Pooled} = \frac{\sum_{b=1}^{h} w_b \tau_{10,b}}{\sum_{b=1}^{h} w_b}, \tag{17}$$

where $w_b$ is the block weight. We set $w_b = n_b p_{10,b}$, so that blocks are weighted by their complier sample populations, but other options exist, such as $w_b = 1$.

Consider OLS estimation of the following extension of (9) to blocked RCTs:

$$y_{ib} = \sum_{b=1}^{h} \alpha_b S_{ib} + \sum_{b=1}^{h} \tau_{ITT,b} S_{ib} \tilde{T}_{ib} + \tilde{\mathbf{x}}_{ib} \boldsymbol{\beta} + \eta_{ib}, \tag{18}$$

where $\tilde{T}_{ib} = T_{ib} - p_b$ and $\tilde{\mathbf{x}}_{ib} = \mathbf{x}_{ib} - \bar{\mathbf{x}}_b$ are block-centered variables; $\bar{\mathbf{x}}_b = \frac{1}{n_b}\sum_{i=1}^{n_b} \mathbf{x}_{ib}$ are covariate means; and $\eta_{ib}$ is the error term. The OLS estimator for $\tau_{ITT,b}$ in (18) is,

$$\hat{\tau}_{ITT,b} = (\bar{y}_b^1 - \bar{y}_b^0) - (\bar{\mathbf{x}}_b^1 - \bar{\mathbf{x}}_b^0)\hat{\boldsymbol{\beta}}, \tag{19}$$

where $\bar{y}_b^t$ and $\bar{\mathbf{x}}_b^t$ are observed treatment and control group means. Similarly, using the treatment receipt indicator, $d_{ib}$, as the dependent variable in (18) yields the OLS estimator, $\hat{\pi}_{ITT,b} = (\bar{d}_b^1 - \bar{d}_b^0) - (\bar{\mathbf{x}}_b^1 - \bar{\mathbf{x}}_b^0)\hat{\boldsymbol{\gamma}}$. Thus, a consistent LATE estimator for block $b$ is, $\hat{\tau}_{10,b} = \hat{\tau}_{ITT,b}/\hat{\pi}_{ITT,b}$.

Because a mini-experiment is conducted in each block, we obtain the following CLT for $\hat{\tau}_{10,b}$ that applies Theorem 1 to each block:

**Theorem 3.** Assume (*C1*)-(*C10*) in Theorem 1 for each $b \in (1, 2, \ldots, h)$, where treatment assignments are mutually independent across blocks, and the following added condition:

(C7a) The block shares, $q_b = n_b/n \to q_b^*$ as $n \to \infty$, where $q_b^* > 0$ and $\sum_{b=1}^{h} q_b^* = 1$.

Then, as $n \to \infty$ for fixed $h$, the LATE estimator, $\hat{\tau}_{10,b}$, is a consistent estimator for $\tau_{10,b}$, and

$$\frac{\hat{\tau}_{10,b} - \tau_{10,b}}{\sqrt{\text{Var}(\bar{Q}_b)}} \xrightarrow{d} N(0,1),$$

where $\text{Var}(\bar{Q}_b)$ is defined as in (13) at the block level using the residual, $R_{ib}(t) = \frac{1}{\pi_{ITT,b}}(\tilde{Y}_{ib}(t) - \tilde{\mathbf{x}}_{ib}\boldsymbol{\beta} - \tau_{10,b}(\tilde{D}_{ib}(t) - \tilde{\mathbf{x}}_{ib}\boldsymbol{\gamma}))$.

The proof of this theorem follows the proof of Theorem 1 for each block using $R_{ib}(t)$. Note that (C7a) ensures that the block sizes, $n_b$, grow with $n$. A variance estimator, $\widehat{\text{Var}}(\bar{Q}_b)$, can be obtained using (16) for each block, where $s_{R_b}^2(t)$ is calculated using block residuals with denominators, $(n_b^1 - Vq_b p_b - 1)$ and $(n_b^0 - Vq_b(1 - p_b) - 1)$.

Next, we provide a CLT for the pooled estimator across blocks, $\hat{\tau}_{10,Pooled} = \frac{\sum_{b=1}^{h} n_b p_{10,b} \hat{\tau}_{10,b}}{\sum_{b=1}^{h} n_b p_{10,b}}$.

**Corollary 1.** Assume the conditions of Theorem 3 and that $p_{10,b}$ is known. Then, as $n \to \infty$ for fixed $h$, the pooled LATE estimator, $\hat{\tau}_{10,Pooled}$, is consistent for $\tau_{10,Pooled}$ in (17), and

$$\frac{\hat{\tau}_{10,Pooled} - \tau_{10,Pooled}}{\sqrt{\text{Var}(\bar{Q}_{Pooled})}} \xrightarrow{d} N(0,1),$$

where $Var(\bar{Q}_{Pooled}) = \frac{\sum_{b=1}^{h}(n_b p_{10,b})^2 \text{Var}(\bar{Q}_b)}{(\sum_{b=1}^{h} n_b p_{10,b})^2}$.

This CLT follows because the $\hat{\tau}_{10,b}$ estimates are asymptotically independent by applying Corollary 1 in Schochet et al. (2022). We can estimate $\text{Var}(\bar{Q}_{Pooled})$ using $\widehat{\text{Var}}(\bar{Q}_b)$ and $\hat{p}_{10,b}$ for each block. Hypothesis testing can be conducted using t-tests with $df = (n - V - 2h)$.

### 4.2. Clustered RCTs

In clustered RCTs, groups (e.g., schools, hospitals, or communities) are randomized rather than individuals. Consider a clustered, non-blocked RCT with $m$ total clusters, where $m^1 = mp$ are assigned to the treatment group and $m^0 = m(1-p)$ are assigned to the control group. All persons in the same cluster have the same treatment assignment. We index clusters by $j$. Thus, $T_j = 1$ for treatment clusters and 0 for control clusters, $n_j$ is the number of persons in cluster $j$, $Y_{ij}(t)$ is the potential outcome for person $i$ in cluster $j$, and so on. We assume individual-level data are available for analysis, but our results also apply to data averaged to the cluster level.

As discussed in Schochet and Chang (2011), compliance for clustered RCTs can pertain to *both* clusters who offer the treatment, $B_j(T_j)$, and individuals within clusters who receive the treatment, $D_{ij}(T_j) = D_{ij}(T_j, B_j(T_j))$. In this setting, potential outcomes, $Y_{ij}(B_j(T_j), D_{ij}(T_j))$, are now a function of both $B_j$ and $D_{ij}$, and there are 16 possible principal strata defined by all combinations of $(B_j(1), D_j(1), B_j(0), D_j(0))$ in the two research conditions.

In this setting, Schochet and Chang (2011) specify multilevel SUTVA, monotonicity, and exclusion restrictions that identify the following LATE estimand for the (1,1,0,0) stratum that pertains to complier individuals in complier clusters:

$$\tau_{1100} = \frac{\tau_{ITT,c}}{p_{1100}} = \frac{\tau_{ITT,c}}{\pi_{ITT,c}}. \tag{20}$$

Here, $\tau_{ITT,c} = \bar{\bar{Y}}(1) - \bar{\bar{Y}}(0)$ is the ITT estimand on the potential outcome; $\bar{\bar{Y}}(t) = \frac{1}{\sum_{j=1}^m w_j} \sum_{j=1}^m w_j \bar{Y}_j(t)$ is the weighted mean potential outcome for $t \in \{1,0\}$; $\bar{Y}_j(t) = \frac{1}{n_j} \sum_{i=1}^{n_j} Y_{ij}(t)$ is the cluster mean; $w_j$ is the cluster weight (e.g., $w_j = n_j$ or 1); $\bar{\bar{D}}(t) = \frac{1}{\sum_{j=1}^m w_j} \sum_{j=1}^m w_j \bar{D}_j(t)$ and $\bar{D}_j(t) = \frac{1}{n_j} \sum_{i=1}^{n_j} D_{ij}(t)$ are potential treatment receipt rates; $p_{1100,c} > 0$ is the sample share of (1,1,0,0) compliers; and $\pi_{ITT,c} = \bar{\bar{D}}(1) - \bar{\bar{D}}(0)$ is the ITT estimand on treatment receipt.

With these identifying conditions, we can estimate $\tau_{1100}$ using similar methods as for the non-clustered RCT by generalizing the working regression model in (9) as follows:

$$y_{ij} = \alpha_c + \tau_{ITT,c}\tilde{T}_j + \tilde{\mathbf{x}}_{ij}\boldsymbol{\beta}_c + e_{i,c}, \tag{21}$$

where $\tilde{\mathbf{x}}_{ij} = (\mathbf{x}_{ij} - \bar{\bar{\mathbf{x}}})$ are centered covariates; $\bar{\bar{\mathbf{x}}} = \frac{1}{\sum_{j=1}^m w_j}\sum_{j=1}^m w_j \bar{\mathbf{x}}_j$ and $\bar{\mathbf{x}}_j = \frac{1}{n_j}\sum_{i=1}^{n_j} \mathbf{x}_{ij}$ are covariate means; $e_{i,c}$ is the error; and $\alpha_c$, $\tau_{ITT,c}$, and $\boldsymbol{\beta}_c$ are parameters. A parallel model applies to the treatment receipt model using $d_{ij}$ as the dependent variable. Estimating these two models using OLS or weighted least squares (WLS) yields the following consistent LATE estimator:

$$\hat{\tau}_{1100} = \frac{\hat{\tau}_{ITT,c}}{\hat{\pi}_{ITT,c}} = \frac{(\bar{\bar{y}}^1 - \bar{\bar{y}}^0) - (\bar{\bar{\mathbf{x}}}^1 - \bar{\bar{\mathbf{x}}}^0)\hat{\boldsymbol{\beta}}_c}{(\bar{\bar{d}}^1 - \bar{\bar{d}}^0) - (\bar{\bar{\mathbf{x}}}^1 - \bar{\bar{\mathbf{x}}}^0)\hat{\boldsymbol{\gamma}}_c}, \tag{22}$$

where $\hat{\boldsymbol{\beta}}_c$ and $\hat{\boldsymbol{\gamma}}_c$ are parameter estimates; $\bar{\bar{y}}^t = \frac{1}{m^t \bar{w}^t}\sum_{j:T_j=1}^m w_j \bar{y}_j$ and $\bar{y}_j = \frac{1}{n_j}\sum_{i=1}^{n_j} y_{ij}$ are observed outcome means, $\bar{w}^t = \frac{1}{m^t}\sum_{j:T_j=t}^m w_j$ is the mean cluster weight, and similarly for $\bar{\bar{d}}^t$ and $\bar{\bar{\mathbf{x}}}^t$. Note that $\bar{\bar{y}}^t$, $\bar{\bar{d}}^t$, and $\bar{\bar{\mathbf{x}}}^t$ are themselves ratio estimators because $\bar{w}^1$ and $\bar{w}^0$ are random variables due to cluster-level randomization (if the weights differ across clusters).

Theorem 4 in Appendix C proves a CLT for $\hat{\tau}_{1100}$ as $m \to \infty$, where the proof uses similar methods as in Theorem 1 above and results in Schochet et al. (2022) who allow for general weights. To state the theorem, let $\bar{R}_j(t) = \left(\frac{1}{\pi_{ITT,c}}\right)\left(\frac{w_j}{\bar{w}}\right)\left[\tilde{\bar{Y}}_j(t) - \tilde{\bar{\mathbf{x}}}_j\boldsymbol{\beta}_c - \tau_{10,c}\left(\tilde{\bar{D}}_j(t) - \tilde{\bar{\mathbf{x}}}_j\boldsymbol{\gamma}_c\right)\right]$ denote weighted cluster residuals with variance $S_R^2(t) = \frac{1}{m}\sum_{i=1}^m \bar{R}_j^2$, where $\bar{w} = \frac{1}{m}\sum_{j=1}^m w_j$ is the mean cluster weight, $\tilde{\bar{Y}}_j(t) = \bar{Y}_j(t) - \bar{\bar{Y}}(t)$ is the centered potential outcome, and similarly for $\tilde{\bar{\mathbf{x}}}_j$ and $\tilde{\bar{D}}_j(t)$. Further, let $S^2(\tau_{1100}) = \frac{1}{m-1}\sum_{j=1}^m (\bar{R}_j(1) - \bar{R}_j(0))^2$ denote the heterogeneity term.

**Theorem 4**. Under the regularity conditions presented in Appendix C, as $m \to \infty$,

$$\frac{\hat{\tau}_{1100} - \tau_{1100}}{\sqrt{\text{Var}(\bar{Q}_c)}} \xrightarrow{d} N(0,1),$$

where

$$\text{Var}(\bar{Q}_c) = \frac{S_{\tilde{R}}^2(1)}{m^1} + \frac{S_{\tilde{R}}^2(0)}{m^0} - \frac{S^2(\tau_{1100})}{m}. \tag{23}$$

Intuitively, the CLT for the clustered RCT generalizes the CLT for the non-clustered RCT by using cluster averages. The main addition to the regularity conditions is a bound on the variance of the weights as $m$ grows so that a weak law of large numbers can be applied to $\bar{w}^t$.

A conservative, plug-in estimator for the variance in (23) based on estimated cluster-level residuals that generalizes the variance estimator in (16) for the non-clustered design is:

$$\hat{\text{Var}}(\bar{Q}_c) = \frac{s_{\tilde{R}}^2(1)}{m^1} + \frac{s_{\tilde{R}}^2(0)}{m^0}, \tag{24}$$

where

$$s_{\tilde{R}}^2(t) = \frac{1}{\hat{\pi}_{ITT,c}^2 (m^t - k_c^t - 1)} \sum_{j:T_j=t}^{m} \frac{w_j^2}{(\bar{w}^t)^2} \left( \tilde{\bar{y}}_j^t - \tilde{\bar{\mathbf{x}}}_j^t \hat{\boldsymbol{\beta}}_c - \hat{\tau}_{10,c} (\tilde{\bar{d}}_j^t - \tilde{\bar{\mathbf{x}}}_j^t \hat{\boldsymbol{\gamma}}_c) \right)^2$$

are residuals for group $t \in \{1,0\}$; $\tilde{\bar{y}}_j^t = \bar{y}_j - \bar{\bar{y}}^t$, $\tilde{\bar{d}}_j^t = \bar{d}_j - \bar{\bar{d}}^t$, and $\tilde{\bar{\mathbf{x}}}_j^t = \bar{\mathbf{x}}_j - \bar{\bar{\mathbf{x}}}^t$ are group-centered means; and $k_c^1 = Vp$ and $k_c^0 = V(1-p)$ are *df* corrections. Hypothesis testing can be performed using t-tests with $df = (m - V - 2)$ that is based on $m$ not $n$, noting that $df$ issues for clustered designs are complex (Donald and Lang, 2007; Cameron et al., 2015).

Finally, we can extend our analysis for the clustered RCT to the blocked, clustered RCT using the same methods as in Section 4.1 for the non-clustered RCT, where we obtain a LATE estimator for each block, and then weight those to obtain a pooled estimator across blocks.

## 5. Simulation analysis

We conducted simulations to examine the finite sample statistical properties of our design-based LATE estimators under the assumed identifying conditions and strong instruments, focusing on the non-clustered RCT. For the simulations, we applied the variance estimator in (16) for models with and without a baseline covariate. We also used a variant of (16) where we subtracted $\frac{1}{n}(s_R(1) - s_R(0))^2$, a lower bound on the heterogeneity term in (13) based on the Cauchy-Schwarz inequality (Aronow et al., 2014 discuss sharper bounds). Finally, we used the standard IV variance estimator with constant treatment effects using ivreg in R.

*5.1. Simulation setup*

We ran simulations for total samples of $n = 200$ and $400$ and a treatment assignment rate of $p = .5$ ($.4$ or $.6$ for some runs). To allow for correlations between potential treatment receipt decisions and outcomes, we adapted the binary latent variable approach of Vytlacil (2002) and Hazard and Lowe (2022). First, we generated the random variable, $\delta_i \sim N(0,1)$, to measure the latent tendency of person $i$ to receive the treatment. Next, because under monotonicity the sample shares of always-takers and never-takers are $p_{11} = \bar{D}(0)$ and $p_{00} = 1 - \bar{D}(1)$, we used $\delta_i$ to generate data for $D_i(0)$ and $D_i(1)$ using the following binary choice probit model for given $p_{11}$ and $p_{00}$: $D_i(0) = \mathbb{I}(\delta_i \leq \Phi^{-1}(p_{11}))$ and $D_i(1) = \mathbb{I}(\delta_i \leq \Phi^{-1}(1 - p_{00}))$, where $\mathbb{I}$ is the indicator function and $\Phi$ is the standard normal cumulative distribution function.

We then generated potential outcomes in the control condition using, $Y_i(0) = \varphi \delta_i + \eta_i$, where $\eta_i \sim N(0,1)$ are random errors and $\varphi$ is a scalar that determines $\rho_{\delta, Y0} = Corr(\delta_i, Y_i(0))$. We also generated a baseline covariate using, $x_i = Y_i(0) + u_i$, where $u_i \sim N(0, \sigma_u^2)$ are random errors, with $\sigma_u^2$ selected to achieve a given $R_{Y0,X}^2$ value for a regression of $Y_i(0)$ on $x_i$.

Next, we divided the $n$ individuals into compliers, always-takers, and never-takers based on their simulated $D_i(0)$ and $D_i(1)$ values. We then generated potential outcomes in the treatment condition for those in principal stratum $(d, d') \in \{(1,0), (1,1), (0,0)\}$ as follows:

$$Y_{i,10}(1) = Y_{i,10}(0) + \theta_{i,10};$$
$$Y_{i,11}(1) = Y_{i,11}(0); \text{ and } Y_{i,00}(1) = Y_{i,00}(0);$$
(25)

where $\theta_{i,10} = \psi \delta_i + \nu_i$ capture treatment heterogeneity for compliers; $\nu_i \sim N(0, \sigma_\nu^2)$ are errors; and $\psi$ and $\sigma_\nu^2$ were calculated to yield pre-set values of $\rho_{\delta,\theta} = Corr(\delta_i, \theta_{i,10})$ and $\sigma_\theta^2$.

We generated 5 simulated datasets to help guard against unusual draws and report average results. For each dataset, we conducted 10,000 replications, randomly assigning units to either the treatment or control group. For each replication, we set $y_i = Y_i(T_i)$ and $d_i = D_i(T_i)$, estimated the ITT regression models, and stored the LATE and variance estimates.

We ran separate simulations for $p_{11} = \bar{D}(0) = .2$, and $.4$, and for $\bar{D}(1) - \bar{D}(0) = .3$ and $.5$ to ensure $T_i$ is a strong instrument for $d_i$ (see Section 3.2). For all runs, we set $\rho_{\delta,Y0} = .3$, $R^2_{Y0,X} = .4$, $\rho_{\delta,\theta} = .1$, and $\sigma^2_\theta = \frac{1}{3}Var(Y_i(0)) = \frac{1}{3}(\varphi^2 + 1)$, which leads to larger outcome variances in the treatment than control condition.

*5.2. Simulation results*

Table 1 and Appendix Table D.1 present the simulation results. We find negligible biases for all specifications with and without baseline covariates. In contrast, biases of LATE estimators based on one-stage OLS regressions of $y_i$ on $d_i$ are much larger, ranging from -.489 to -.240 across the Table 1 specifications (not shown). Confidence interval coverage is close to 95 percent using t-distribution cutoff values, but with some over-coverage. Smaller sample sizes and smaller compliance effects yield more over-coverage. These results suggest that relatively large samples are required to rely on asymptotic results for design-based LATE analyses.

Estimated standard errors are close to "true" values as measured by the standard deviation of the LATE estimates across replications. As predicted by the theory in Section 3.4, the design-based and IV variances are very similar, where the IV variance is slightly larger if $p = .6$ and slightly smaller if $p = .4$, which occurs because variances were constructed to be larger in the treatment than control condition in our simulations (Table D.1). Finally, subtracting the bound on the heterogeneity term does not change the overall findings or improve performance.

## 6. Empirical application using the NYC voucher experiment

To demonstrate our design-based LATE estimators, we used baseline and outcome data from the NYC School Choice Scholarships Foundation Program (SCSF) (Mayer et al., 2002). SCSF was funded by philanthropists to provide scholarships to public school students in grades kindergarten to 4 from low-income families to attend any participating NYC private school. In spring 1997, more than 20,000 students applied to receive a voucher. SCSF then used random lotteries to offer 3-year vouchers of up to $1,400 annually to 1,000 eligible families in the treatment group. Of the remaining families not offered the voucher, 960 were randomly selected to the control group.

SCSF assisted the treatment group in finding private-school placements. About 80 percent of treatment families used a voucher within the first follow-up year, of whom 98 percent attended parochial schools. In addition, 7 percent of controls also attended a private school without a voucher. Thus, the LATE estimand is pertinent for this study due to treatment noncompliance. Further, the finite-population framework is realistic as the study included only a very small percentage of NYC families who applied for a scholarship. Thus, the study results cannot be generalized to a broader voucher program that would involve all children in NYC or elsewhere.

We estimate LATE effects for the full sample and two race/ethnicity subgroups as defined in Mayer et al. (2002): African Americans and Latinos who each comprise about 47 percent of the sample. The study authors hypothesized that African Americans might benefit more from the vouchers as they tended to live in more disadvantaged communities with lower-performing public schools. Private school attendance rates were slightly higher for African Americans than Latinos for the treatment group (82 versus 77 percent) but not for the control group.

Following Mayer et al. (2002), the primary outcomes for our analysis are composite national percentile rankings in math and reading from the study-administered Iowa Test of Basic Skills (ITBS). We focus on first follow-up year test scores with a response rate of 78 and 71 percent for treatments and controls. Our goal is not to replicate study results but to illustrate our estimators.

The voucher study was a blocked RCT. Applicants from schools with average test scores below the city median were assigned a higher probability of winning a scholarship, and blocks were also formed by lottery date and family size (with 30 blocks in total). The design is also partly clustered because families were randomized, where all eligible children within a family could receive a scholarship; 30 percent of families had at least two children in the evaluation.

We used (18) for LATE estimation and (16) for variance estimation for each block, where blocks were weighted by their complier sizes ($w_b = n_b \hat{p}_{10,b}$) to obtain overall LATE effects. To adjust for clustering, we averaged data to the family level. Following the original study, we used weights to adjust for missing follow-up test scores. We ran models without covariates and those that included baseline ITBS scores to increase precision, though they were not collected for kindergarteners. Following the original study, other demographic covariates were not included in the models due to the large number of blocks. Finally, we also estimated an IV model with block fixed effects (but not block-by-treatment interactions) as used in the original study.

Table 2 presents the ITT and LATE findings that mirror those in Mayer et al. (2002) and Krueger and Zhu (2004). We find that the offer of a voucher had no effect on test scores overall or for Latinos across specifications. The effects on African Americans are also not significant at the 5 percent level for the model without baseline test scores. However, these effects become positive and significant for the model with baseline scores (that excludes the kindergarteners), where standard errors are reduced by 12 percent. These ITT and LATE effects are 4.7 and 6.2 percentile ranking points, which translate into .26 and .34 standard deviation increases, with a significant F-test for the subgroup interaction ($p$-value = .033). A detailed reanalysis of the original study data, however, cautions that the results for African Americans are sensitive to alternative samples and race/ethnicity definitions and should be interpreted carefully (Krueger and Zhu 2004). Finally, consistent with the literature, the design-based standard errors are almost entirely due to the ITT standard errors (i.e., the correction terms in (14) matter little).

## 7. Conclusions

This article considered design-based methods for estimation of the LATE estimand for RCTs with treatment noncompliance using simple least squares methods. Our ratio estimators derive from the Neyman-Rubin-Holland model that underlies experiments combined with standard monotonicity and exclusion restrictions found in the IV literature. The design-based approach is appealing in that it applies to continuous, binary, and discrete outcomes, allows for heterogeneous treatment effects, and is nonparametric in that it makes no assumptions about the distribution of potential outcomes or the model functional form.

We developed a finite-population CLT for the LATE estimator under the non-clustered RCT, allowing for baseline covariates to improve precision, and extended the unified theory to more general blocked and clustered RCTs commonly used in practice. We provided simple consistent variance estimators using regression residuals from the outcome and treatment receipt models. Our re-analysis of the NYC voucher experiment demonstrated the simplicity of the methods, while maintaining statistical rigor.

Simulations for the non-clustered RCT suggest that the design-based LATE estimators yield low bias and confidence interval coverage near nominal levels, although with some over-coverage. The simulations suggest also that LATE analyses require relatively large sample sizes to justify the use of asymptotic results. A future research topic is to conduct a more complete simulation analysis for the full range of RCT designs, including clustered and blocked RCTs, and to examine statistical performance under violations to the underlying assumptions.

Our theory and simulations both point to the similarity of the design-based and commonly-used IV variance estimators for LATE analyses. Thus, the distinction between the finite- and superpopulation perspectives for LATE estimation is likely to be blurred in practice. Nonetheless, study protocols should specify the adopted framework to help interpret the LATE (and ITT) findings, as the finite-population perspective assumes LATE results pertain to compliers in the study sample only rather than to a representative superpopulation of compliers who would be subject to broader implementation of the treatment.

The free *RCT-YES* software ([www.rct-yes.com](www.rct-yes.com)), funded by the U.S. Department of Education, estimates LATE effects for both full sample and baseline subgroup analyses using the design-based methods discussed in this article using either R or Stata. The software allows for general weights and also allows for multi-armed trials with multiple treatment conditions.

**Data Availability Statement**. The NYC Voucher data for the empirical analysis were obtained under a restricted data use license agreement with Mathematica. Per license requirements, these data cannot be shared with journal readers. However, to the best of my knowledge, these data can be obtained, and I would be happy to provide the SAS and R programs used for the analysis.
**Funding.** This research did not receive any specific grant from funding agencies in the public, commercial, or not-for-profit sectors.
**Conflict of Interest.** The author has no conflicts of interest associated with this article.
**Acknowledgements**. None.

**Table 1.** Simulation results for the LATE estimators

| Model specification and LATE variance estimator | Bias of LATE estimator[a] | Confidence interval coverage | True standard error[a,b] | Mean estimated standard error |
|---|---|---|---|---|
| **Model without covariates** | | | | |
| Sample: $n = 400$; $\bar{D}(0) = .2$; $\bar{D}(1) = .5$ | | | | |
| Design-based (DB) | .012 | .962 | .355 | .357 |
| DB with heterogeneity term | .012 | .962 | .355 | .357 |
| Instrumental variables (IV) | .012 | .962 | .355 | .357 |
| Sample: $n = 400$; $\bar{D}(0) = .2$; $\bar{D}(1) = .7$ | | | | |
| DB | .005 | .958 | .226 | .225 |
| DB with heterogeneity term | .005 | .958 | .226 | .225 |
| IV | .005 | .958 | .226 | .225 |
| Sample: $n = 200$; $\bar{D}(0) = .2$; $\bar{D}(1) = .5$ | | | | |
| DB | .025 | .968 | .508 | .512 |
| DB with heterogeneity term | .025 | .967 | .508 | .511 |
| IV | .025 | .968 | .508 | .512 |
| Sample: $n = 200$; $\bar{D}(0) = .2$; $\bar{D}(1) = .7$ | | | | |
| DB | .008 | .960 | .323 | .326 |
| DB with heterogeneity term | .008 | .960 | .323 | .326 |
| IV | .008 | .960 | .323 | .326 |
| **Model with one covariate** | | | | |
| Sample: $n = 400$; $\bar{D}(0) = .2$; $\bar{D}(1) = .5$ | | | | |
| DB | .007 | .964 | .282 | .285 |
| DB with heterogeneity term | .007 | .964 | .282 | .285 |
| IV | .007 | .964 | .282 | .286 |
| Sample: $n = 400$; $\bar{D}(0) = .2$; $\bar{D}(1) = .7$ | | | | |
| DB | .002 | .960 | .187 | .186 |
| DB with heterogeneity term | .002 | .959 | .187 | .186 |
| IV | .002 | .960 | .187 | .186 |
| Sample: $n = 200$; $\bar{D}(0) = .2$; $\bar{D}(1) = .5$ | | | | |
| DB | .018 | .970 | .422 | .424 |
| DB with heterogeneity term | .018 | .969 | .422 | .423 |
| IV | .018 | .970 | .422 | .426 |
| Sample: $n = 200$; $\bar{D}(0) = .2$; $\bar{D}(1) = .7$ | | | | |
| DB | .005 | .961 | .261 | .264 |
| DB with heterogeneity term | .005 | .961 | .261 | .264 |
| IV | .005 | .962 | .261 | .265 |

Notes. See text for simulation details. The calculations assume a treatment assignment rate of $p = .5$, and normally distributed outcomes and covariates. For each specification, the figures are based on 10,000 simulations for each of 5 potential outcome draws, and the findings average across the 5 draws. LATE estimates are based on the estimator in (12), design-based standard errors are obtained using (16), and the bound on the heterogeneity term is discussed in the text. IV estimates are obtained using ivreg in R.

LATE = Local average treatment effect; DB = Design-based; IV = Instrumental variables.

[a] Biases and true standard errors are the same for all specifications within each sample size category because they use the same data and model for LATE estimation.

[b] True standard errors are measured as the standard deviation of the estimated treatment effects across simulations.

**Table 2.** Estimated effects on composite test scores for the NYC voucher experiment

| Model specification | Overall sample | African American | Latino |
|---|---|---|---|
| **Model excludes baseline test scores** | | | |
| Design-based ITT estimator | 0.25 | 2.54 | -0.86 |
| | (1.06) | (1.45) | (1.58) |
| Design-based LATE estimator | 0.35 | 3.40 | -1.19 |
| | (1.46) | (1.95) | (2.20) |
| Instrumental variables LATE estimator with block fixed effects | -0.02 | 2.94 | -1.44 |
| | (1.62) | (2.18) | (2.51) |
| **Model includes baseline test scores** | | | |
| Design-based ITT estimator | 1.70 | 4.70* | 0.50 |
| | (1.01) | (1.27) | (1.44) |
| Design-based LATE estimator | 2.30 | 6.15* | 0.66 |
| | (1.37) | (1.71) | (1.92) |
| Instrumental variables LATE estimator with block fixed effects | 2.16 | 6.08* | 0.47 |
| | (1.40) | (1.78) | (2.19) |
| Student sample size (without / with baseline test scores) | 2,012 / 1,434 | 902 / 643 | 964 / 682 |

Notes: Standard errors are in parentheses. Estimates are measured in percentile ranking points and are weighted to adjust for follow-up test score nonresponse.

ITT = Intention-to-treat; LATE = Local average treatment effect.

* Statistically significant at the 5 percent level, two-tailed test.

**Appendix: Theorem Proofs and Additional Simulation Results**

**A. Proof of Theorem 1**

To prove Theorem 1, we first prove consistency and then the central limit theorem (CLT). We use definitions and notation from the main text. Note that the proof also applies to the model without covariates by setting $\boldsymbol{\beta} = \mathbf{0}$ and/or $\boldsymbol{\gamma} = \mathbf{0}$. Our proof uses results in Li and Ding (2017) and Scott and Wu (1981).

*A.1. Proof of consistency*

Our goal is to show that $\hat{\tau}_{10} - \tau_{10} \xrightarrow{p} 0$ as $n \to \infty$. To do this, because $\hat{\tau}_{10} = \hat{\tau}_{ITT}/\hat{\pi}_{ITT}$, we first prove consistency of $\hat{\tau}_{ITT}$ and $\hat{\pi}_{ITT}$, and then apply the continuous mapping theorem. Thus, we proceed by first showing that $\hat{\tau}_{ITT} \xrightarrow{p} \mu_Y(1) - \mu_Y(0)$, where $\mu_Y(1)$ and $\mu_Y(0)$ are finite limiting values of $\bar{Y}(1)$ and $\bar{Y}(0)$, the mean potential outcomes in the treatment and control conditions. We then show that $\hat{\pi}_{ITT} \xrightarrow{p} p_{10}^* = \mu_D(1) - \mu_D(0) > 0$, where $\mu_D(1)$ and $\mu_D(0)$ are finite limiting values of $\bar{D}(1)$ and $\bar{D}(0)$.

To show that $\hat{\tau}_{ITT} \xrightarrow{p} \mu_Y(1) - \mu_Y(0)$, consider ordinary least squares (OLS) estimation of the outcome model in (9) in the main text, where $\tilde{\mathbf{z}}_i = (1 \ \tilde{T}_i, \tilde{\mathbf{x}}_i)$ is the vector of model explanatory variables with parameter vector, $\boldsymbol{\theta}$. The OLS estimator for $\boldsymbol{\theta}$ is,

$$\hat{\boldsymbol{\theta}} = \begin{pmatrix} \hat{\alpha} \\ \hat{\tau}_{ITT} \\ \hat{\boldsymbol{\beta}} \end{pmatrix} = (\frac{1}{n}\sum_{i=1}^{n}\tilde{\mathbf{z}}_i'\tilde{\mathbf{z}}_i)^{-1}\frac{1}{n}\sum_{i=1}^{n}\tilde{\mathbf{z}}_i'y_i, \qquad (A.1)$$

where $y_i = T_i Y_i(1) + (1 - T_i)Y_i(0)$ is the observed outcome. Expanding this estimator yields,

$$\begin{pmatrix} \hat{\alpha} \\ \hat{\tau}_{ITT} \\ \hat{\boldsymbol{\beta}} \end{pmatrix} = \begin{bmatrix} 1 & 0 & 0 \\ 0 & p(1-p) & \frac{1}{n}\sum_i \tilde{T}_i \tilde{\mathbf{x}}_i \\ 0 & \frac{1}{n}\sum_i \tilde{\mathbf{x}}_i' \tilde{T}_i & \frac{1}{n}\sum_i \tilde{\mathbf{x}}_i' \tilde{\mathbf{x}}_i \end{bmatrix}^{-1} \begin{bmatrix} \frac{1}{n}\sum_i y_i \\ \frac{1}{n}\sum_i \tilde{T}_i y_i \\ \frac{1}{n}\sum_i \tilde{\mathbf{x}}_i' y_i \end{bmatrix}. \tag{A.2}$$

As $n \to \infty$, we have in (A.2) that $\frac{1}{n}\sum_{i=1}^n \tilde{\mathbf{z}}_i' \tilde{\mathbf{z}}_i$ converges to a block diagonal matrix because,

$$\frac{1}{n}\sum_i \tilde{T}_i \tilde{\mathbf{x}}_i = p(1-p)[\bar{\mathbf{x}}^1 - \bar{\mathbf{x}}^0] \xrightarrow{p} p^*(1-p^*)[\boldsymbol{\mu}_X - \boldsymbol{\mu}_X] = 0, \tag{A.3}$$

where $\boldsymbol{\mu}_X$ is the limiting value of $\bar{\mathbf{x}}$, assumed finite by (*C10*). This result invokes the weak law of large numbers (WLLN) for each covariate using Theorem B in Scott and Wu (1981), which applies because the covariate variances, $S_{x,v}^2$, are assumed to have finite limits by (*C10*). The WLLN implies that $\bar{\mathbf{x}}^t - \bar{\mathbf{x}} \xrightarrow{p} 0$, which further implies that $(\bar{\mathbf{x}}^t - \boldsymbol{\mu}_X) = (\bar{\mathbf{x}}^t - \bar{\mathbf{x}}) - (\bar{\mathbf{x}} - \boldsymbol{\mu}_X) \xrightarrow{p} 0$, so that $\bar{\mathbf{x}}^t \xrightarrow{p} \boldsymbol{\mu}_X$. The result in (A.3) also relies on (*C7*) that $p$ converges to $p^*$ and the continuous mapping theorem.

Because the design matrix is block diagonal in the limit, it follows that $\hat{\tau}_{ITT}$ has the same limiting value as $\frac{1}{np(1-p)}\sum_{i=1}^n \tilde{T}_i y_i$. To calculate this limit, we use the following result:

$$\frac{1}{np(1-p)}\sum_{i=1}^n \tilde{T}_i y_i = \bar{y}^1 - \bar{y}^0 \xrightarrow{p} \mu_Y(1) - \mu_Y(0), \tag{A.4}$$

which holds using a similar argument as for (A.3) where we can again apply the WLLN using Theorem B in Scott and Wu (1981), which yields, $\bar{y}^t - \bar{Y}(t) \xrightarrow{p} 0$ and $(\bar{y}^t - \mu_Y(t)) = (\bar{y}^t - \bar{Y}(t)) - (\bar{Y}(t) - \mu_Y(t)) \xrightarrow{p} 0$. Thus,

$$\hat{\tau}_{ITT} \xrightarrow{p} \mu_Y(1) - \mu_Y(0), \tag{A.5}$$

which establishes consistency.

For the CLT proof, we will also need asymptotic values for the OLS estimates of the intercept, $\hat{\alpha}$, and covariate parameter vector, $\hat{\boldsymbol{\beta}}$. First, note that $\hat{\alpha}$ has the same limiting value as $\bar{y} = \frac{1}{n}\sum_i y_i$, so we can use a similar argument as for $\hat{\tau}_{ITT}$ to show that,

$$\hat{\alpha} \xrightarrow{p} p^*\mu_Y(1) + (1-p^*)\mu_Y(0) = \alpha^*, \tag{A.6}$$

which is the limiting value of the mean potential outcome.

Similarly, the OLS estimator for the covariates, $\hat{\boldsymbol{\beta}}$, has the same limiting value as $\left[\frac{1}{n}\sum_{i=1}^{n} \tilde{\mathbf{x}}_i'\tilde{\mathbf{x}}_i\right]^{-1}\left[\frac{1}{n}\sum_{i=1}^{n} \tilde{\mathbf{x}}_i' y_i\right]$. For the first bracketed term, we have that,

$$\frac{1}{n}\sum_{i=1}^{n} \tilde{\mathbf{x}}_i'\tilde{\mathbf{x}}_i \xrightarrow{p} \boldsymbol{\Omega}_{\mathbf{x}}^2, \tag{A.7}$$

where $\boldsymbol{\Omega}_{\mathbf{x}}^2$ is the limiting value of $\mathbf{S}_{\mathbf{x}}^2$, which is finite by (*C10*). For the second bracketed term, if we insert the relation, $y_i = T_i Y_i(1) + (1-T_i)Y_i(0)$, we find that,

$$\frac{1}{n}\sum_{i=1}^{n} \tilde{\mathbf{x}}_i' y_i = \frac{1}{n}\sum_{i=1}^{n} T_i \tilde{\mathbf{x}}_i' Y_i(1) + \frac{1}{n}\sum_{i=1}^{n}(1-T_i)\tilde{\mathbf{x}}_i' Y_i(0). \tag{A.8}$$

We can now invoke the WLLN for each term on the right-hand side of (A.8), again using Theorem B in Scott and Wu (1981). We can apply this theorem because our conditions imply that $\mathbf{S}_{\mathbf{x}Y}^2(t)$ has a finite limit (which is not specified directly). To see this, we have for each covariate $v$ that,

$$\begin{aligned}
S_{x_v Y}^2(t) &\leq \frac{1}{n}\sum_{i=1}^{n} \tilde{x}_{i,v}^2 Y_i^2(t) \leq \max_{1\leq i\leq n}\{\tilde{x}_{i,v}^2\}\frac{1}{(n-1)}\sum_{i=1}^{n} Y_i^2(t) \\
&= \max_{1\leq i\leq n}\{\tilde{x}_{i,v}^2\}[S_Y^2(t) - \frac{n}{(n-1)}\bar{Y}^2(t)]\pi_{ITT}^2.
\end{aligned} \tag{A.9}$$

Thus, because of (*C9*) and the assumptions that $S_{x,v}^2$, $S_Y^2(t)$, $\pi_{ITT}$, and $n^t/n$ each have finite limits, we have that $S_{x_v Y}^2(t) = O_p(1)$, which establishes that $\mathbf{S}_{\mathbf{x}Y}^2(t)$ has a finite limit.

We find then that the first term in (A.8) has the same asymptotic value as $p\mathbf{S}^2_{\mathbf{x},Y}(1)$ and the second term has the same asymptotic value as $(1-p)\mathbf{S}^2_{\mathbf{x},Y}(0)$. Thus, if $\mathbf{\Omega}^2_{\mathbf{x},Y}(t)$ is the finite limiting value of $\mathbf{S}^2_{\mathbf{x},Y}(t)$, we find that,

$$\frac{1}{n}\sum_{i=1}^n \tilde{\mathbf{x}}'_i y_i \xrightarrow{p} p^*\mathbf{\Omega}^2_{\mathbf{x},Y}(1) + (1-p^*)\mathbf{\Omega}^2_{\mathbf{x},Y}(0). \tag{A.10}$$

Finally, putting together the pieces in (A.7) and (A.10), we find that as $n \to \infty$,

$$\hat{\boldsymbol{\beta}} \xrightarrow{p} (\mathbf{\Omega}^2_{\mathbf{x}})^{-1}\left[p^*\mathbf{\Omega}^2_{\mathbf{x},Y}(1) + (1-p^*)\mathbf{\Omega}^2_{\mathbf{x},Y}(0)\right] = \boldsymbol{\beta}^*, \tag{A.11}$$

which are the asymptotic regression coefficients that would be obtained from a hypothetical regression of $[pY_i(1) + (1-p)Y_i(0)]$ on $\tilde{\mathbf{x}}_i$.

The same approach can be used to prove consistency of the OLS estimates in the treatment receipt model. This leads to the following results: $\hat{\pi}_{ITT} \xrightarrow{p} p^*_{10} = \mu_D(1) - \mu_D(0)$, $\hat{\alpha}_D \xrightarrow{p} p\mu_D(1) + (1-p)\mu_D(0)$, and $\hat{\boldsymbol{\gamma}} \xrightarrow{p} (\mathbf{\Omega}^2_{\mathbf{x}})^{-1}\left[p^*\mathbf{\Omega}^2_{\mathbf{x},D}(1) + (1-p^*)\mathbf{\Omega}^2_{\mathbf{x},D}(0)\right]$, where $\hat{\alpha}_D$ is the estimated intercept and $\mathbf{\Omega}^2_{\mathbf{x},D}(t)$ is the finite limiting covariance matrix for $\mathbf{S}^2_{\mathbf{x},D}(t)$. Note that $S^2_D(t)$, $S^2_{Y,D}(t)$, and $\mathbf{S}^2_{\mathbf{x},D}(t)$ have finite limits by (C8) and (C10) because $D_i(t)$ is binary, and thus, these finite limits are not specified directly in (C10). For example, $S^2_{Y,D}(t) = \bar{D}(t)(1-\bar{D}(t))(\bar{Y}_1(t) - \bar{Y}_0(t))$ is a function of first moments only, where $\bar{Y}_1(t)$ and $\bar{Y}_0(t)$ are the means of $Y_i(t)$ for those with $\bar{D}(t) = 1$ and $\bar{D}(t) = 0$, and similarly for $S^2_{Y,D}(t)$ and $\mathbf{S}^2_{\mathbf{x},D}(t)$.

Finally, because $\hat{\tau}_{ITT} - \tau_{ITT} \xrightarrow{p} 0$, $\hat{\pi}_{ITT} - \pi_{ITT} \xrightarrow{p} 0$, and $\lim_{n\to\infty} \pi_{ITT} = p^*_{10} > 0$ by (C8), we have by the continuous mapping theorem that $\hat{\tau}_{10} - \tau_{10} \xrightarrow{p} 0$. ∎

*A.2. Proof of CLT*

To prove the CLT, we express $(\hat{\tau}_{10} - \tau_{10})$ as follows:

$$(\hat{\tau}_{10} - \tau_{10}) = \left[\frac{\hat{\tau}_{ITT} - \tau_{10}\hat{\pi}_{ITT}}{\pi_{ITT}}\right]\left[\frac{\pi_{ITT}}{\hat{\pi}_{ITT}}\right], \quad (A.12)$$

where $\hat{\tau}_{ITT} = [(\bar{y}^1 - \bar{y}^0) - (\bar{\mathbf{x}}^1 - \bar{\mathbf{x}}^0)\hat{\boldsymbol{\beta}}]$ and $\hat{\pi}_{ITT} = [(\bar{d}^1 - \bar{d}^0) - (\bar{\mathbf{x}}^1 - \bar{\mathbf{x}}^0)\hat{\boldsymbol{\gamma}}]$ are estimated ITT effects, recalling that $\pi_{ITT} > 0$ by (C8). We now examine each square bracketed term in turn, recalling that $R_i(t) = \frac{1}{\pi_{ITT}}(\tilde{Y}_i(t) - \tilde{\mathbf{x}}_i\boldsymbol{\beta} - \tau_{10}(\tilde{D}_i(t) - \tilde{\mathbf{x}}_i\boldsymbol{\gamma}))$ is the residualized, centered potential outcome in the treatment or control condition, with variance, $S_R^2(t) = \frac{1}{(n-1)}\sum_{i=1}^n R_i^2$.

For the first bracketed term in (A.12), we have that,

$$\left[\frac{\hat{\tau}_{ITT} - \tau_{10}\hat{\pi}_{ITT}}{\pi_{ITT}}\right] = \frac{1}{n^1}\sum_{i:T_i=1}^n \hat{R}_i(1) - \frac{1}{n^0}\sum_{i:T_i=0}^n \hat{R}_i(0) = \bar{r}(1) - \bar{r}(0), \quad (A.13)$$

where $\hat{R}_i(t) = \frac{1}{\pi_{ITT}}\left(\tilde{Y}_i(t) - \tilde{\mathbf{x}}_i\hat{\boldsymbol{\beta}} - \tau_{10}(\tilde{D}_i(t) - \tilde{\mathbf{x}}_i\hat{\boldsymbol{\gamma}})\right)$ is based on the estimated regression parameters, $\hat{\boldsymbol{\beta}}$ and $\hat{\boldsymbol{\gamma}}$. Note that the first equality uses the relation, $(\bar{Y}(1) - \bar{Y}(0)) - \tau_{10}(\bar{D}(1) - \bar{D}(0)) = 0$, which holds by definition because $\tau_{10} = \frac{\bar{Y}(1) - \bar{Y}(0)}{\bar{D}(1) - \bar{D}(0)}$.

Under our theorem conditions, we have from Theorem 1 in Li and Ding (2017) that,

$$\frac{\bar{r}(t)}{\sqrt{\frac{(1-f^t)}{n^t}S_R^2(t)}} \xrightarrow{d} N(0,1), \quad (A.14)$$

for $t \in \{1,0\}$, where $f^t = n^t/n$ and $S_R^2(t)$ have finite limits by assumption. Further, we have from Theorem 4 in Li and Ding (2017) that,

$$\frac{\bar{r}(1) - \bar{r}(0)}{\sqrt{Var(\bar{Q})}} \xrightarrow{d} N(0,1), \quad (A.15)$$

where $Var(\bar{Q})$ is defined as in (13) in the main text.

Turning to the second bracketed term in (A.12), we have from our consistency result in online appendix A.1 that $\pi_{ITT}/\hat{\pi}_{ITT} \xrightarrow{p} 1$ as $n \to \infty$. Thus, the CLT in Theorem 1 follows using the CLT in (A.15) and Slutsky's theorem. ∎

## B. Proof of Theorem 2 on the consistency of $s_R^2(t)$

Our goal is to prove that $s_R^2(t) - S_R^2(t) \xrightarrow{p} 0$ as $n \to \infty$ for $t \in \{1,0\}$, where

$$s_R^2(t) = \frac{1}{\hat{\pi}_{ITT}^2(n^t - k^t - 1)} \sum_{i:T_i=t}^n \left(\tilde{y}_i^t - \tilde{\mathbf{x}}_i^t \hat{\boldsymbol{\beta}} - \hat{\tau}_{10}(\tilde{d}_i^t - \tilde{\mathbf{x}}_i^t \hat{\boldsymbol{\gamma}})\right)^2$$

and

$$S_R^2(t) = \frac{1}{\pi_{ITT}^2(n-1)} \sum_{i=1}^n \left(\tilde{Y}_i(t) - \tilde{\mathbf{x}}_i \boldsymbol{\beta} - \tau_{10}(\tilde{D}_i(t) - \tilde{\mathbf{x}}_i \boldsymbol{\gamma})\right)^2.$$

To proceed, we expand $s_R^2(t)$ using the decomposition in (14) in the main text, which yields,

$s_R^2(t) = s_{RY}^2(t) + s_{RD}^2(t) + s_{RYD}^2(t)$, where $s_{RY}^2(t) = \frac{1}{\hat{\pi}_{ITT}^2(n^t-k^t-1)} \sum_{i:T_i=t}^n (\tilde{y}_i^t - \tilde{\mathbf{x}}_i \hat{\boldsymbol{\beta}})^2$, $s_{RD}^2(t) = \frac{\hat{\tau}_{10}^2}{\hat{\pi}_{ITT}^2(n^t-k^t-1)} \sum_{i:T_i=t}^n (\tilde{d}_i^t - \tilde{\mathbf{x}}_i \hat{\boldsymbol{\gamma}})^2$, and $s_{RYD}^2(t) = \frac{-2\hat{\tau}_{10}}{\hat{\pi}_{ITT}^2(n^t-k^t-1)} \sum_{i:T_i=t}^n (\tilde{y}_i^t - \tilde{\mathbf{x}}_i \hat{\boldsymbol{\beta}})(\tilde{d}_i^t - \tilde{\mathbf{x}}_i \hat{\boldsymbol{\gamma}})$.

Consider the $s_{RY}^2(t)$ component, where the same approach applies to $s_{RD}^2(t)$ and $s_{RYD}^2(t)$. To show that $s_{RY}^2(t) - S_{RY}^2(t) \xrightarrow{p} 0$, we adapt Lemma A.3 in Li and Ding (2017) to our context. First, it is useful to define $s_{RY}^{*2}(t) = \frac{1}{\pi_{ITT}^2(n^t-1)} \sum_{i:T_i=t}^n (\tilde{y}_i^t - \tilde{\mathbf{x}}_i \hat{\boldsymbol{\beta}})^2$, so that $s_{RY}^2(t) = \left[\frac{\pi_{ITT}^2}{\hat{\pi}_{ITT}^2}\right]\left[\frac{n^t-1}{n^t-k^t-1}\right] s_{RY}^{*2}(t)$. Because $\frac{\pi_{ITT}^2}{\hat{\pi}_{ITT}^2} \xrightarrow{p} 1$ as shown in online appendix A.1 and $\frac{n^t-1}{n^t-k^t-1}$ also converges to 1, the continuous mapping theorem implies that $s_{RY}^{*2}(t)$ and $s_{RY}^2(t)$ have the same limiting value. Thus, we focus on $s_{RY}^{*2}(t)$, which also helps to reduce notation.

Next, we expand $s_{RY}^{*2}(t)$ as follows:

$$s_{RY}^{*2}(t) = s_Y^{*2}(t) - 2\mathbf{s}_{\mathbf{x},Y}^{*2}(t)\hat{\boldsymbol{\beta}} + \hat{\boldsymbol{\beta}}' \mathbf{s}_{\mathbf{x}}^{*2}(t)\hat{\boldsymbol{\beta}}, \tag{B.1}$$

where $s_Y^{*2}(t) = \frac{1}{\pi_{ITT}^2 (n^t-1)} \sum_{i:T_i=t}^{n} (Y_i(t) - \bar{y}^t)^2$, $\mathbf{s}_{\mathbf{x},Y}^{*2}(t) = \frac{1}{\pi_{ITT}^2 (n^t-1)} \sum_{i:T_i=t}^{n} \tilde{\mathbf{x}}_i'(Y_i(t) - \bar{y}^t)$, and

$\mathbf{s}_{\mathbf{x}}^{*2}(t) = \frac{1}{\pi_{ITT}^2 (n^t-1)} \sum_{i:T_i=t}^{n} \tilde{\mathbf{x}}_i' \tilde{\mathbf{x}}_i$. From online appendix A.1, we showed that $\widehat{\boldsymbol{\beta}} - \boldsymbol{\beta} \xrightarrow{p} \mathbf{0}$, so we proceed by examining the limiting behavior of the estimated covariances in (B.1).

Consider $\mathbf{s}_{\mathbf{x},Y}^{*2}(t)$. Focusing on covariate $v$, we can express $s_{x_v,Y}^{*2}(t)$ as,

$$s_{x_v,Y}^{*2}(t) = \frac{1}{\pi_{ITT}^2 (n^t - 1)} \sum_{i:T_i=t}^{n} \tilde{x}_{i,v}(Y_i(t) - \bar{Y}(t))$$
$$- \frac{n^t}{\pi_{ITT}^2 (n^t - 1)} (\bar{x}_v^t - \bar{x}_v)(\bar{y}^t - \bar{Y}(t)). \tag{B.2}$$

For the first term in (B.2), using Markov's (or Chebyshev's) inequality, we find that,

$$\frac{1}{n^t} \sum_{i:T_i=t}^{n} \tilde{x}_{i,v}(Y_i(t) - \bar{Y}(t)) - \frac{1}{n} \sum_{i=1}^{n} \tilde{x}_{i,v}(Y_i(t) - \bar{Y}(t)) = O_p\left(\sqrt{\frac{S_{x_v Y}^2(t)}{n^t}}\right) = o_p(1). \tag{B.3}$$

The final equality holds because $S_{x_v Y}^2(t)$ has a finite limit using the same arguments as in (A.9) in online appendix A.1.

For the second term in (B.2), we can again use Markov's inequality to show that,

$(\bar{x}_v^t - \bar{x}_v) = O_p\left(\sqrt{S_{x,v}^2(t)/n^t}\right) = O_p(1/n) = o_p(1)$, which holds because $S_{x,v}^2(t)$ has a finite limit by (C10) and $n^t/n = O_p(1)$ by (C7). Similarly, $(\bar{y}^t - \bar{Y}(t)) = O_p\left(\sqrt{S_Y^2(t)/n^t}\right) = o_p(1)$.

Finally, because $n^t/[\pi_{ITT}^2 (n^t - 1)] = O_p(1)$, we see that the second term in (B.2) is $o_p(1)$.

Thus, inserting (B.3) into (B.2) and subtracting $S_{x_v,Y}^2(t)$ yields,

$$s_{x_v,Y}^{*2}(t) - S_{x_v,Y}^2(t) = o_p(1) + \frac{n^t(n-1)}{(n^t - 1)n} S_{x_v,Y}^2(t) - S_{x_v,Y}^2(t). \tag{B.4}$$

If we combine the last two terms in (B.4), we find that, $\left(1 - \frac{n^t(n-1)}{(n^t-1)n}\right) S_{x_v,Y}^2(t) = o_p(1)$. Thus,

$s_{x_v,Y}^{*2}(t) - S_{x_v,Y}^2(t) \xrightarrow{p} 0$ for each covariate $v$, and hence, $\mathbf{s}_{\mathbf{x},Y}^{*2}(t) - \mathbf{S}_{\mathbf{x},Y}^2(t) \xrightarrow{p} \mathbf{0}$.

The same proof with parallel conditions applies to $s_Y^{*2}(t)$ and $\mathbf{s}_\mathbf{x}^{*2}(t)$ in (B.1) as well as to the other covariances, $s_{RD}^{*2}(t)$ and $s_{RYD}^{*2}(t)$, that comprise $s_R^{*2}(t)$. Pulling these results together and using the continuous mapping theorem establishes that $s_R^2(t) - S_R^2(t) \xrightarrow{p} 0$. ∎

## C. Theorem 4 for the clustered RCT

The LATE estimator for the clustered RCT is,

$$\hat{\tau}_{1100} = \frac{\hat{\tau}_{ITT,c}}{\hat{\pi}_{ITT,c}} = \frac{(\bar{y}^1 - \bar{y}^0) - (\bar{\mathbf{x}}^1 - \bar{\mathbf{x}}^0)\widehat{\boldsymbol{\beta}}_c}{(\bar{\bar{d}}^1 - \bar{\bar{d}}^0) - (\bar{\mathbf{x}}^1 - \bar{\mathbf{x}}^0)\widehat{\boldsymbol{\gamma}}_c}, \tag{C.1}$$

which is a ratio of the OLS (or WLS) ITT estimators for the outcome and treatment receipt indicator. In this section, we present a CLT for $\hat{\tau}_{1100}$ with a proof that closely follows the methods in online appendix A and results in Schochet et al. (2022).

Our CLT relies on several assumptions drawn from the literature that generalize (*C1*)-(*C5*) for the non-clustered RCT to the clustered RCT setting. First, we require cluster randomization:

*(C2a): Complete randomization of clusters*: For fixed $m^1$, if $\mathbf{t}_{clus} = (t_1, \ldots, t_m)$ is any vector of randomization realizations such that $\sum_{j=1}^m t_{clus} = m^1$, then $Prob(\mathbf{T}_{clus} = \mathbf{t}_{clus}) = \binom{m}{m^1}^{-1}$.

Second, we invoke the (*S1*) to (*S5*) multilevel SUTVA, monotonicity, and exclusion restrictions in Schochet and Chang (2011) for the clustered RCT, that we label as (*C1a*) and (*C3a*)-(*C6a*) to be consistent with our notation. As discussed in the main text, these conditions allow for compliance decisions to be made by *both* clusters who offer the treatment and individuals within clusters who receive the treatment, and identify the LATE estimand for the (1,1,0,0) population that pertain to complier individuals in complier clusters. Third, we invoke a condition on the existence of individuals in the (1,1,0,0) principal stratum:

*(C7a): Presence of multilevel compliers*: $0 < p_{1100} < 1$.

Before presenting our theorem, we first provide several definitions that pertain to weighted cluster-level averages and parallel those for the non-clustered RCT. First, recall that $\bar{R}_j(t) = \left(\frac{1}{\pi_{ITT,c}}\right)\left(\frac{w_j}{\bar{w}}\right)\left[\tilde{\bar{Y}}_j(t) - \tilde{\bar{x}}_j\boldsymbol{\beta}_c - \tau_{10,c}\left(\bar{\tilde{D}}_j(t) - \tilde{\bar{x}}_j\boldsymbol{\gamma}_c\right)\right]$ is the weighted cluster-level residual with variance, $S_{\bar{R}}^2(t) = \frac{1}{m-1}\sum_{j=1}^m \bar{R}_j^2(t)$, and covariance, $S_{\bar{R}}^2(1,0)$. Further, recall that $S^2(\tau_{1100}) = \frac{1}{m-1}\sum_{j=1}^m (\bar{R}_j(1) - \bar{R}_j(0))^2$ is the heterogeneity term. We also require cluster-level variances for the outcomes; $S_{\bar{Y}}^2(t) = \frac{1}{(m-1)\pi_{ITT,c}^2}\sum_{j=1}^m \frac{w_j^2}{\bar{w}^2}\tilde{\bar{Y}}_j^2(t)$; for each covariate $v \in \{1, \dots, V\}$, $S_{\bar{x},v}^2 = \frac{1}{(m-1)\pi_{ITT,c}^2}\sum_{j=1}^m \frac{w_j^2}{\bar{w}^2}\tilde{\bar{x}}_{j,v}^2$; and for the full covariate set, $\mathbf{S}_{\bar{\mathbf{x}}}^2 = \frac{1}{m}\sum_{j=1}^m \frac{w_j^2}{\bar{w}^2}\tilde{\bar{\mathbf{x}}}_j'\tilde{\bar{\mathbf{x}}}_j$. We also need several cluster-level covariances between $\tilde{\bar{Y}}_j(t)$, $\tilde{\bar{\mathbf{x}}}_j$, and $\bar{D}_j(t)$: $\mathbf{S}_{\bar{\mathbf{x}},\bar{Y}}^2(t) = \frac{1}{m}\sum_{j=1}^m \frac{w_j^2}{\bar{w}^2}\tilde{\bar{\mathbf{x}}}_j'\bar{Y}_j(t)$; $\mathbf{S}_{\bar{\mathbf{x}}\bar{Y}}^2(t) = \frac{1}{m}\sum_{j=1}^m \frac{w_j^2}{\bar{w}^2}\left(\tilde{\bar{\mathbf{x}}}_j'\bar{Y}_j(t) - \bar{\boldsymbol{\theta}}\right)^2$, where $\bar{\boldsymbol{\theta}} = \frac{1}{m}\sum_{j=1}^m \frac{w_j^2}{\bar{w}^2}\tilde{\bar{\mathbf{x}}}_j'\bar{Y}_j(t)$; and similarly for $\mathbf{S}_{\bar{\mathbf{x}},\bar{D}}^2(t)$, $\mathbf{S}_{\bar{\mathbf{x}}\bar{D}}^2(t)$, $S_{\bar{Y},\bar{D}}^2(t)$, and $S_{\bar{Y}\bar{D}}^2(t)$. Finally, let the variance of the weights be, $S^2(w) = \frac{1}{m-1}\sum_{j=1}^m (w_j - \bar{w})^2$.

We now present our theorem.

**Theorem 4.** Assume (*C1a*) to (*C7a*), and the following conditions for $t \in \{1,0\}$:

(*C8a*) Letting $g(t) = \max_{1 \leq j \leq m}\{\bar{R}_j^2(t)\}$, as $m \to \infty$,

$$\frac{1}{(m^t)^2}\frac{g(t)}{\text{Var}(\bar{Q}_c)} \to 0.$$

(*C9a*) $f^1 = m^1/m$ and $f^0 = m^0/m$ have limiting values, $p^*$ and $(1 - p^*)$, for $0 < p^* < 1$.

(*C10a*) The complier share, $p_{1100} = \pi_{ITT,c}$, converges to $p_{1100}^*$, for $0 < p_{1100}^* < 1$, and similarly for the other existent principal stratum specified in Schochet and Chang (2011), where all the $p^*$ values sum to 1.

(*C11a*) As $m \to \infty$,
$$\frac{(1 - f^t)S^2(w)}{m^t\bar{w}^2} \to 0.$$

(*C12a*) Letting $h_v(t) = \max_{1 \le j \le m}\{\tilde{\tilde{x}}_{j,v}^2\}$ for all $v \in \{1, \ldots, V\}$, as $m \to \infty$,

$$\frac{1}{\min(m^1, m^0)} \frac{h_v(t)}{S_{\tilde{x},v}^2} \to 0.$$

(*C13a*) $S_{\hat{R}}^2(t), S_{\hat{R}}^2(1,0), S_{\bar{Y}}^2, S_{\bar{D}}^2, S_{\tilde{x},v}^2, \mathbf{S}_{\bar{\mathbf{x}}}^2, \mathbf{S}_{\bar{\mathbf{x}},\bar{Y}}^2(t), S_{\bar{D}\bar{Y}}^2$, and $\mathbf{S}_{\bar{\mathbf{x}},\bar{D}}^2$ have finite limits.

Then, as $m \to \infty$, the IV estimator, $\hat{\tau}_{1100}$, is a consistent estimator for $\tau_{1100}$, and

$$\frac{\hat{\tau}_{1100} - \tau_{1100}}{\sqrt{\mathrm{Var}(\bar{Q}_c)}} \xrightarrow{d} N(0,1),$$

where $\mathrm{Var}(\bar{Q}_c)$ is defined as in (23) in the main text.

*Proof.* To establish consistency, we first apply Theorem 1 in Schochet et al. (2022) who show that $\hat{\tau}_{ITT,c} - \tau_{ITT,c} \xrightarrow{p} 0$. We can apply this result because the conditions in Theorem 3 that pertain to the outcomes and covariates parallel those in Theorem 1 in Schochet et al. (2022). Further, because Theorem 3 adds parallel conditions for the treatment receipt indicator, the same methods can be used to prove that $\hat{\pi}_{ITT,c} - \pi_{ITT,c} \xrightarrow{p} 0$. Thus, combining these results, we find that $\hat{\tau}_{1100} = \hat{\tau}_{ITT,c}/\hat{\pi}_{ITT,c}$ is a consistent estimator for $\tau_{1100}$ by the continuous mapping theorem. Thus, $\hat{\tau}_{1100} - \tau_{1100} \xrightarrow{p} 0$.

Next, to prove the CLT result in Theorem 4, we follow a similar approach as for the proof of Theorem 1 in online appendix A. First, we express $(\hat{\tau}_{1100} - \tau_{1100})$ as follows:

$$(\hat{\tau}_{1100} - \tau_{1100}) = \left[\frac{\hat{\tau}_{ITT,c} - \tau_{1100}\hat{\pi}_{ITT,c}}{\pi_{ITT,c}}\right]\left[\frac{\pi_{ITT,c}}{\hat{\pi}_{ITT,c}}\right], \tag{C.2}$$

where $\hat{\tau}_{ITT,c} = \left[(\bar{y}^1 - \bar{y}^0) - (\bar{\mathbf{x}}^1 - \bar{\mathbf{x}}^0)\hat{\boldsymbol{\beta}}_c\right]$ and $\hat{\pi}_{ITT,c} = \left[(\bar{d}^1 - \bar{d}^0) - (\bar{\mathbf{x}}^1 - \bar{\mathbf{x}}^0)\hat{\boldsymbol{\gamma}}_c\right]$ are the estimated ITT effects using least squares. For the first bracketed term in (C.2), we have that,

$$\left[\frac{\hat{\tau}_{ITT,c} - \tau_{1100}\hat{\pi}_{ITT,c}}{\pi_{ITT,c}}\right] = \frac{1}{m^1}\sum_{j:T_j=1}^{m} \hat{\bar{R}}_j(1) - \frac{1}{m^0}\sum_{j:T_j=0}^{m} \hat{\bar{R}}_j(0) = \bar{r}(1) - \bar{r}(0), \tag{C.3}$$

where the cluster residual, $\hat{\bar{R}}_j(t) = \left(\frac{1}{\pi_{ITT,c}}\right)\left(\frac{w_j}{\bar{w}}\right)[\tilde{\bar{Y}}_j(t) - \tilde{\bar{\mathbf{x}}}_j\hat{\boldsymbol{\beta}}_c - \tau_{10,c}(\tilde{\bar{D}}_j(t) - \tilde{\bar{\mathbf{x}}}_j\hat{\boldsymbol{\gamma}}_c)]$, is based on the estimated regression parameters, $\hat{\boldsymbol{\beta}}_c$ and $\hat{\boldsymbol{\gamma}}_c$. Note that the first equality uses the relation, $(\bar{\bar{Y}}(1) - \bar{\bar{Y}}(0)) - \tau_{1100}(\bar{\bar{D}}(1) - \bar{\bar{D}}(0)) = 0$, which holds by definition under our identification assumptions because $\tau_{1100} = \frac{\bar{\bar{Y}}(1) - \bar{\bar{Y}}(0)}{\bar{\bar{D}}(1) - \bar{\bar{D}}(0)}$.

Under the Theorem 4 conditions, we have from Theorem 1 in Schochet et al. (2022) that,

$$\frac{\bar{\bar{r}}(1) - \bar{\bar{r}}(0)}{\sqrt{Var(\bar{Q}_c)}} \xrightarrow{d} N(0,1), \tag{C.4}$$

where $Var(\bar{Q}_c)$ is defined as in (23) in the main text. Thus, using (C.3) and (C.4), we have that,

$$\frac{1}{\sqrt{Var(\bar{Q}_c)}}\left[\frac{\hat{\tau}_{ITT,c} - \tau_{1100}\hat{\pi}_{ITT,c}}{\pi_{ITT,c}}\right] \xrightarrow{d} N(0,1). \tag{C.5}$$

Finally, for the second bracketed term in (C.2), we have from our consistency result above that $\frac{\pi_{ITT,c}}{\hat{\pi}_{ITT,c}} \xrightarrow{p} 1$ as $m \to \infty$. Thus, the CLT in Theorem 4 follows using the CLT result in (C.5) and Slutsky's theorem. ∎

## D. Additional simulation results

Table D.1 below displays additional simulation results for the alternative specifications discussed in the main text.

**Table D.1.** Additional simulation results for the LATE estimators

| LATE estimator | Bias of LATE estimator[a] | Confidence interval coverage | True standard error[a,b] | Mean estimated standard error |
|---|---|---|---|---|
| **Model with one covariate** | | | | |
| Sample: $n = 400$; $p = .5$; $\bar{D}(0) = .4$; $\bar{D}(1) = .7$ | | | | |
| DB | .009 | .965 | .304 | .308 |
| DB with heterogeneity term | .009 | .965 | .304 | .308 |
| IV | .009 | .965 | .304 | .309 |
| Sample: $n = 400$; $p = .5$; $\bar{D}(0) = .4$; $\bar{D}(1) = .9$ | | | | |
| DB | .003 | .957 | .176 | .175 |
| DB with heterogeneity term | .003 | .957 | .176 | .174 |
| IV | .003 | .957 | .176 | .175 |
| Sample: $n = 200$; $p = .5$; $\bar{D}(0) = .4$; $\bar{D}(1) = .7$ | | | | |
| Design-based (DB) | .025 | .972 | .444 | .438 |
| DB with heterogeneity term | .025 | .971 | .444 | .437 |
| Instrumental variables (IV) | .025 | .972 | .444 | .439 |
| Sample: $n = 200$; $p = .5$; $\bar{D}(0) = .4$; $\bar{D}(1) = .9$ | | | | |
| DB | .003 | .961 | .236 | .239 |
| DB with heterogeneity term | .003 | .961 | .236 | .238 |
| IV | .003 | .962 | .236 | .239 |
| Sample: $n = 400$; $p = .4$; $\bar{D}(0) = .2$; $\bar{D}(1) = .5$ | | | | |
| DB | .008 | .963 | .305 | .307 |
| DB with heterogeneity term | .008 | .962 | .305 | .307 |
| IV | .008 | .960 | .305 | .303 |
| Sample: $n = 400$; $p = .4$; $\bar{D}(0) = .2$; $\bar{D}(1) = .7$ | | | | |
| DB | .002 | .960 | .188 | .186 |
| DB with heterogeneity term | .002 | .960 | .188 | .185 |
| IV | .002 | .956 | .188 | .182 |
| Sample: $n = 400$; $p = .6$; $\bar{D}(0) = .2$; $\bar{D}(1) = .5$ | | | | |
| DB | .008 | .963 | .334 | .324 |
| DB with heterogeneity term | .008 | .962 | .334 | .324 |
| IV | .008 | .966 | .334 | .329 |
| Sample: $n = 400$; $p = .6$; $\bar{D}(0) = .2$; $\bar{D}(1) = .7$ | | | | |
| DB | .003 | .958 | .181 | .184 |
| DB with heterogeneity term | .003 | .958 | .181 | .183 |
| IV | .003 | .963 | .181 | .189 |

Notes. See text for simulation details. The calculations assume normally distributed outcomes and covariates. For each specification, the figures are based on 10,000 simulations for each of 5 potential outcome draws, and the findings average across the 5 draws. LATE estimates are based on the estimator in (12), design-based standard errors are obtained using (16), and the bound on the heterogeneity term is discussed in the text. IV estimates are obtained using ivreg in R.

LATE = Local average treatment effect; DB = Design-based; IV = Instrumental variables.

[a] Biases and true standard errors are the same for all specifications within each sample size category because they use the same data and model for LATE estimation.

[b] True standard errors are measured as the standard deviation of the estimated treatment effects across simulations.